\documentclass[useAMS,usenatbib]{mn2e}
\usepackage{array,etoolbox}
\usepackage{graphicx}
\usepackage{subfig}
\usepackage{color}
\usepackage{float}
\newcommand{\be}{\begin{equation}}
\newcommand{\ee}{\end{equation}}

\title[]{  Evolutionary outcomes  for pairs of planets  undergoing    orbital migration
and circularization:  second order resonances   and  observed   period ratios   in Kepler's planetary systems }
\author[M. Xiang-Gruess and J.C.B. Papaloizou]{M. Xiang-Gruess$^{1}$ \thanks{E-mail:
mxianggruess@mpifr-bonn.mpg.de } and J. C. B. Papaloizou$^{2}$ \\
$^{1}$ Max-Planck Institut f\"ur Radioastronomie, Auf dem H\"ugel 69, D-53121 Bonn, Germany \\
$^{2}$ DAMTP, University of Cambridge, Wilberforce Road, Cambridge CB3 0WA, UK\\
}
\begin{document}

\date{Accepted . Received ; }

\pagerange{\pageref{firstpage}--\pageref{lastpage}} \pubyear{2002}

\maketitle

\label{firstpage}

\begin{abstract}\

In order to study the  origin of the architectures of low mass planetary systems,
we   perform  numerical  surveys of the evolution of  pairs  of coplanar planets in the mass range $(1-4)\ \rmn{M}_{\oplus}.$ 
These  evolve for up to $2\times10^7 \rmn{yr}$  under a range of orbital  migration torques  and circularization
rates assumed to arise through  interaction with a protoplanetary disc. 

 Near the inner disc boundary, significant variations of viscosity, interaction with density waves or with the stellar magnetic field 
could occur and halt migration, but allow circularization to continue. This was  modelled by modifying the migration and circularization rates.

Runs terminated without an extended period of circularization in the absence of migration torques gave rise to either a collision, 
or a system close to a resonance. These were mostly first order with  a few $\%$ terminating in second order resonances.  Both planetary eccentricities were small $< 0.1$ and all
 resonant angles liberated.  
This type of survey produced  only a limited range of period ratios and cannot reproduce Kepler observations.

When circularization alone operates in the final stages, divergent migration occurs causing period ratios to increase. Depending on its strength
 the whole period ratio range between $1$ and $2$ can be obtained. A few systems close to  second order commensurabilities  also occur.  In contrast to when arising  through
  convergent migration, resonant trapping does not occur and  resonant angles circulate. Thus the behaviour of the resonant angles may indicate the form of migration that led to near resonance.

\end{abstract}


\section{Introduction}\label{Intro}

The Kepler  space  telescope has been very successful in revealing  the diversity of the architectures of  exoplanetary systems.
To date, Kepler has detected over 4000 planet candidates \citep[see][]{Bat2013}, from which
about a thousand  are confirmed planets. About one third of Kepler's candidates are associated 
with multiple transiting systems and the vast majority of them are expected to be multi-planetary systems \citep[][]{Lis2012}.  Amongst these,
Super-Earths with orbital periods less than 20 days are  abundant around Sun-like stars. 
It is unlikely that these planets formed at their current locations, but instead they could  have
migrated inwards from significantly larger radii 
\citep[see for example][for a discussion]{Izi2014}.

Due to interaction between the planets and the protoplanetary disc,  torques  that can 
lead  to orbital migration are  expected to occur  as long as the planets are embedded in the disc 
\citep[e.g.][]{Gol2014}.
For planets with different masses,  migration rates  are expected to  differ  leading to a time-dependent evolution of their orbital period ratio.
When the outermost planet is more massive it migrates  through the disc faster leading to convergent migration.
If the inner planet is more massive, as long as its migration stalls at the inner boundary of the disc, convergent migration will then ultimately ensue.

For systems in which the planetary orbits converge, the planets tend to enter and become locked in a mean-motion resonance 
and migrate together subsequently \citep[e.g.][]{Nel2002, Kle2004}.
In the simplest case of nearly circular and coplanar orbits,  first-order resonances are readily  formed 
between two planets undergoing convergent migration.  For these  the orbital period ratio can be expressed as $(p+1)/p$, with $p$ being an integer.

 Kepler's  candidate   multi-planetary systems   show indications of having been affected by resonances.  Notably there are
excesses in the numbers of systems just wide of the 2:1 and 3:2 commensurabilities  \citep[e.g.][]{Steffen2013}.
However, most systems are non resonant with a wide range of period ratios for  consecutive planets   being observed.
This is apparently in conflict with the notion of widespread convergent migration.
However, much of the previous discussion of this issue has been  based on  utilisation of simplified  protoplanetary disc models 
without  consideration of conditions near the  inner boundary where migration can be halted.  
The proper incorporation of such effects is  problematic as it  is likely to require detailed modelling  
of structural features in the protoplanetary disc resulting from, either significant  variations of effective viscosity
if the transition is between regions of varying levels of turbulence, or
 the  interaction  of the disc with the magnetic field of the central star.
 In the latter case there may be a region where the surface
density increases outwards supporting outward propagating density waves.
 In this paper we assess the effects of a transition at an inner disc boundary,
or  the  possible effects of wake-planet or density wave - planet interaction, 
on the inward migration of a pair of planets through simplified
 prescriptions for the orbital migration and circularization  rates
of the planets.

In order to investigate   the outcome of the migration of a pair of  low mass planets  and  the consequent   architectures of low mass planetary systems, we  have performed numerical  $N$-body simulations of a pair of coplanar  migrating low-mass planets
in the mass range $ (1 - 4) \ \rmn{M}_{\oplus}.$
In  the  surveys we conducted the  planets were initialised on circular coplanar orbits with the outer one being at 1 AU or 5 AU
and  the inner one being just outside either the 2:1 commensurability or the 3:2 commensurability.
The orbital evolution was considered for a large  range of  migration and   circularization times  
which  might be supposed to arise  from interacting with the protoplanetary  disc for times up to $2\times 10^7\ \rmn{yr}$.
We suppose the disc to possess an inner cavity where the structure changes on account of   
for example,  rapid changes in disc
viscosity,  or interaction with the  central star.  We suppose that  on account of  structural changes,   inward migration ceases inside the cavity
while investigating   cases   for which   orbital circularization ceases for both planets,   it continues for only the outer planet,  or  it continues  for both planets.  
The last two  cases result in divergent migration for the two planets,  so modelling the effect of   wake-planet, or planet-density wave
interactions \citep[see][for a discussion]{Baruteau2013}.  Such runs  have to be considered if the final range of period ratios is  to contain  those found in the observations. 

With this modelling,  we  find that the runs in our  surveys can have a variety of ends.  These include a collision between the planets,  a configuration in or close to a first order resonance,
and systems occupying a range of period ratios as is found for the observed systems.
 In a relatively  small number of cases,  second order commensurabilities are found.
The  resonant angles are found to librate only when they are formed through convergent migration as expected.   
Note too that results obtained for two planet systems may also be relevant to systems of higher multiplicity if these can be regarded as being  built up sequentially.

Systems exhibiting second order commensurabilities
 are few in number and little studied in the context of migration models.
However,  further consideration   may  indicate that they could reveal information about their origin, so we have considered them in some detail.
From   the  confirmed exoplanets,   we list   
 systems with period ratios that could potentially place them in second-order resonances  in Table \ref{tab:2nd_reso_kepler}.
All of these have  a fractional deviation  from  a commensurability,   $\ < 6\times 10^{-3},$   which might allow them  to be within it ( see Section \ref{Behaviour}  below). 
However, note that  the fractional deviation for Kepler 365 is less than $6 \times 10^{-4}$ and for  Kepler 29 and  Kepler 87 it is less than  $4\times 10^{-5}.$     

The discovery of  two planets either in or close to
7:5  resonance in HD 41248 was announced by  \citet {Jen2013}. This subsequently became controversial resulting in some uncertainty as to their status
\citep[see][and references therein] {Jen2014}. Note that the innermost   planet  in Kepler 87 is  a giant.
Although the focus of this paper is on the Earth to  Super-Earth   mass range,  this was included for completeness.

\begin{table}
 \begin{tabular}{|c|c|c|}
\hline
Resonance & System& Period ratio \\
\hline
5:3 & Kepler 365 &1.66754  \\
\hline
5:3 & Kepler  262&1.67322 \\
\hline
5:3 & Kepler 87 & 1.66671 \\
\hline
7:5  & HD 41248 & 1.394 $\pm$ 0.005  \\
\hline
9:7 & Kepler 29 &  1.28567   \\
\hline
9:7 & Kepler 417 & 1.29292  \\
\hline
\end{tabular}
\caption{Apart from the case of HD 41248 (see text) , the entries are confirmed  Kepler systems  that are close  to second-order commensurabilities.}
\label{tab:2nd_reso_kepler}
\end{table}

The plan of the paper is as follows. We begin by discussing the orbital  properties of planets of  the mass range of interest  in a second order resonance.
In Sections \ref{sec:theory} - \ref{Behaviour}  we provide a semi-analytic 
 discussion,  giving  expressions for the  eccentricities as a function of
the ratio of circularization time to the migration time  for the case when the planets undergo self-similar migration while maintaining the resonance.
These are found to give results in good agreement with numerical simulations.
In addition  we show that these resonant configurations can occur when the eccentricities are small  provided migration rates are sufficiently small.
We go on to give a brief  discussion of higher order resonances  in Section  \ref{Higherorder}.

In Sections \ref{sec:sim_details} - \ref{initconfig}, we give  the details of the general setup  of the  numerical surveys. We then go on to present the outcomes of the surveys, as outlined  above,  in Sections \ref{results} - \ref{Continuationruns}.
Finally, in Section \ref{sec:conclusion}, we  summarise and discuss our results.

\section{A semi-analytic solution for planets in second-order resonances undergoing migration and orbital circularization} \label{sec:theory}

Here we use the formalism of \citet{Murray1999}. These authors give the equations for the orbital elements of the outer planet adopting a polar coordinate system with origin located  at the  primary mass
as
\begin{eqnarray}
\hspace{-7mm}&&\frac{da_2}{dt}= \frac{2}{n_2a_2}\frac{\partial R}{\partial \lambda_2} , \nonumber \\
\hspace{-7mm}&&\frac{de_2}{dt}= -\frac{\sqrt{1-e_2^2}}{n_2a^2_2e_2}\frac{\partial R}{\partial \varpi_2}    -\frac{\sqrt{1-e^2_2}(1-\sqrt{1-e^2_2})}{n_2a_2^2e_2}\frac{\partial R}{\partial \lambda_2}  , \nonumber \\
\hspace{-7mm}&&\frac{d\varpi_2}{dt}= \frac{\sqrt{1-e^2_2}}{n_2a^2_2e_2}\frac{\partial R}{\partial e_2} , \nonumber \\
\hspace{-7mm}&&\frac{d\lambda_2}{dt}=n_2  -  \frac{2}{n_2a_2}\frac{\partial R}{\partial a_2}    +\frac{\sqrt{1-e^2_2}(1-\sqrt{1-e^2_2})}{n_2a^2_2e_2}\frac{\partial R}{\partial e_2} ,  \label{REM}
\end{eqnarray}
where $a_2$ is the semi-major axis, $e_2$ the eccentricity, $\varpi_2$ the longitude of the periapsis and $\lambda_2$ the mean longitude of the outer planet. 
Here and in the following sections, we  identify  the two planets by applying a subscript   1 to denote the inner planet and  a subscript  2 to denote  the outer planet.
For the outer planet $R= R_D {\rm G} m_1/a_2$, with $m_1$ being the mass of the inner planet and ${\rm G}$ the gravitational constant. Here $R_D$ is the direct part of the disturbing function. 
The indirect part  does not contribute  significant terms in the discussion presented below.
The equations governing the orbital elements of the inner planet take the same form as (\ref{REM})
 but with $a_2, e_2, \varpi_2,$ and $ \lambda_2$ replaced by
$a_1, e_1, \varpi_1,$ and $ \lambda_1$ respectively.   In this case  $R$ is replaced by $ R_D \rmn{G} m_2/a_2. $

The quantity $R_D$ may be developed as a Fourier series in $\lambda_i$ and $\varpi_i,$ $i=1,2.$ 
  In order to discuss  the second order resonance for which
$(p+2)n_2 \sim  pn _1$  where $p$ is an integer $ \ge 3,$       we retain terms containing  $\lambda_1$ and $\lambda_2$ 
 only in the combination $\theta_j = (p+2)\lambda_2-p\lambda_1.$ 
Thus  the retained part  of $R_D$ takes the form 
\begin{eqnarray}
&&R_D=\sum_{n,i }F_{i}^{n}\cos(n\theta_j  -i\varpi_1  +(i-2n)\varpi_2),\label{RD}
 \end{eqnarray}
where $n$ is a positive  integer  or zero, $i$ is a positive or negative integer or zero and  
the amplitudes $F_{i}^{n}$ are functions of $e_1, e_2$ and  $\alpha = a_1/a_2.$
For   second  order resonances,  the  lowest order  amplitudes
  are homogeneous second degree polymomials in the eccentricities
with coefficients that are functions of $\alpha.$
 The  forms of the   $F_{i}^{n}$   can be found  correct to fourth order in the eccentricities
from  \citet{Murray1999}.  See Appendix \ref{MD} below for more explanation.

Note that the terms with $n=0$ in (\ref{RD})
correspond to the standard secular terms  which have to be retained in the discussion of second order resonances considered here.
For second order resonances, the  general angle 

\noindent $n\theta_j -i\varpi_1+(i-2n)\varpi_2$  occurring in (\ref{RD}) 
can be expressed as a linear combination of the two  resonant angles 
\begin{eqnarray}
\hspace{-7mm}&&\phi_1 = (p+2) \lambda_2 - p \lambda_1 - 2 \varpi_1 {\hspace{3mm} \rm and}\label{eq:phi_1} \\
\hspace{-7mm}&& \phi_3 = (p+2) \lambda_2 - p \lambda_1 -  \varpi_1 -\varpi_2  \label{eq:phi_3} 
\end{eqnarray}
 For example
$\phi_2= 2\phi_3-\phi_1 =  (p+2) \lambda_2 - p \lambda_1 - 2 \varpi_2$ and
$\varpi_1 -\varpi_2=\phi_3 -\phi_1.$

Following on from this it can be seen that   a complete set of six  equations for the state variables
$a_i, e_i, i = 1,2,$ and $\phi_1,$ and $\phi_3$ may be obtained 
by taking the first two equations of (\ref{REM}) and their counterparts for planet  $1$, together with equations for $\phi_1$ and 
$\phi_3$ that may be obtained from the third and fourth equations of (\ref{REM}) and their counterparts for planet $1$.

\subsection{Migration and orbital circularization}\label{Migration}

When  a gaseous disc is present,  the interaction of a planet with it  
 will result in changes to its orbital energy and eccentricity.
 Thus if  the orbital energy dissipation rate associated with the interaction of the disc with planet $i$ is $D_i,$
the corresponding rate of change of the semi-major axis is given by 
\be {d a_i\over dt}
=-  {2 a_i^2 \over {\rm G} {\rm M}_* m_i} D_i \label{CEi}\ee
In addition,  tidally induced orbital circularization  causes $e_i$ to change at
a rate given by
 \be \frac{de_i}{dt}  = -e_i/\tau_{c,i}, \label{Ocirc}\ee
where $\tau_{c,i} $ is the circularization time of  planet $i.$
To take account of interaction with the disc, the rates of change given by (\ref{CEi}) and (\ref{Ocirc}) are  added to 
the expressions for $da_i/dt$ and $de_i/dt$  given by the first two equations  of (\ref{REM})
 for planet $2$ and its counterpart for planet $i.$

\subsection{Self-similar migration}\label{s-fmigration}
We consider evolution of the system corresponding to   resonant self-similar  migration in which $a_2/a_1,$ $e_1,$ and $e_2$
are constant,  the angle $\phi_3$ is close to $0$ and  $\phi_1$ is close to $\pi.$
Thus the semi-major axes  decrease on account of migration torques while 
 maintaining the same ratio.  Solutions of this type have been considered by Papaloizou (2003) 
 and Papaloizou \& Szuszkiewicz (2005) for the case of first order resonances and as  the generalisation to consider higher order
 resonances  is straightforward in principle, the reader is referred to those papers for details omitted here.

\subsection{Conservation of angular momentum and energy}\label{conserv}
The total angular momentum of the planetary system
$J=J_1+J_2$ is given by 
\be
 J_1+J_2
 = m_1\sqrt{{\rm G} {\rm M}_* a_1(1-e_1^2)} +  m_2\sqrt{{\rm GM}_* a_2(1-e_2^2)}\ee
and the total  energy $E$  is given by 
\be E = -{{\rm GM}_* m_1\over 2 a_1} -{{\rm GM}_* m_2\over 2 a_2}-m_2R \ee
In the absence of migration and circularization these  
are both conserved. 

 Following Nelson \& Papaloizou (2002), for self-similar  migration induced by interaction with the disc we have  
\begin{eqnarray} 
\hspace{-7mm}&&{dJ\over dt} =J_1{1\over 2  a_1}
{d a_1\over dt}\left( 1 +
{m_2\sqrt{a_2(1-e_2^2)}\over m_1\sqrt{a_1(1-e_1^2)}}\right)=T\nonumber \\
\hspace{-7mm}&& =- \left( \frac{ m_1\sqrt{{\rm GM}_* a_1(1-e_1^2)}}{ \tau_{mig,1}} +\frac{ m_2\sqrt{{\rm GM}_* a_2(1-e_2^2)}}{ \tau_{mig,2 }}  \right), \label{CJ} 
\end{eqnarray}
where the total torque $T = T_1+T_2$ is the sum of the contributions of the 
migration torques acting on $m_1$ and $m_2.$
Here $\tau_{mig,i}$ is defined to be the migration rate of planet $i.$  For an isolated planet with zero eccentricity we have $dn_i/dt= 3n_i/\tau_{mig,i}.$

Neglecting the small contribution  of the interaction energy,  $-m_2R,$ the conservation of total  energy gives
\be {dE\over dt} ={{\rm GM}_* m_1\over 2 a_1^2} {d a_1\over dt}
\left(1+{m_2 a_1\over m_1 a_2}\right)
=-D_1-D_2= - D , \label{CE}\ee
 where the total
 tidally induced orbital energy  loss rate, $D= D_1+D_2,$  is related to
 the torques acting on the planets and their circularization times through 
\be D =  \frac{ {\rm GM}_* m_1 e_1^2}{ a_1(1-e_1^2) \tau_{c,1}}
+\frac{{\rm  GM}_* m_2 e_2^2}{ a_2(1-e_2^2) \tau_{c,2}}
 -\frac{n_2T_2}{ \sqrt{1-e_2^2}} -\frac{n_1T_1}{ \sqrt{1-e_1^2}}
\label{DDD}.\ee

\noindent  By eliminating $d a_1/ dt$ from (\ref{CJ}) and (\ref{CE})
we can obtain a relationship between $e_1, e_2, \tau_c, \tau_{mig,1}$ and $\tau_{mig,2}$ 
which with the help of (\ref{DDD})  becomes
\begin{eqnarray} 
\hspace{-7mm}&& \frac{1}{m_1m_2}\left(\frac {m_1 a_2 e_1^2 }
{\tau_{c,1} (1-e_1^2)} + \frac {m_2a_1 e_2^2 }{\tau_{c,2} (1-e_2^2)}\right)=\nonumber\\
\hspace{-7mm}&&\frac{a_2^{3/2}\sqrt{(1-e_2^2)}-a_1^{3/2}\sqrt{(1-e_1^2)}}
{m_2\sqrt{a_2(1-e_2^2)}+m_1\sqrt{a_1(1-e_1^2)}}
\left( \frac{1}{\tau_{mig,2}}-\frac{1}{\tau_{mig,1}}\right)
\label{ejcons} \end{eqnarray}
We comment that this is general in that it depends only on the conservation laws
and applies for any magnitude of eccentricity or type of resonance. 
Note too that the right hand side of equation (\ref {ejcons})
is proportional to $1/\tau_{mig,2}-1/\tau_{mig,1}$ which is the
difference between the migration rates of the outer and inner planets.
This should be positive, corresponding to convergent migration,
if  this  equation is to be satisfied.
However, self-similar migration with  equilibrium eccentricities was assumed, it may  not apply when
the eccentricities grow continuously.

\subsection{An expression for the eccentricity ratio}\label{eccrat}

We obtain  an equation for the ratio of the eccentricities of the two planets by using (\ref{REM}) to find  an expression
for the rate of change of $\beta=\phi_3 -\phi_1$ which takes the form
\begin{eqnarray}
\hspace{-8mm}&&\frac{d\beta }{dt}=\sum_{n,i }\hspace{-1mm}
\left[\frac{\sqrt{1-e^2_1}m_2}{n_1a^2_1a_2e_1}\frac{\partial F_{i}^{n}}{\partial e_1}
 -\frac{\sqrt{1-e^2_2}m_1}{n_2a^3_2e_2}\frac{\partial F_{i}^{n}}{\partial e_2}\right]\hspace{-1mm}\cos\psi_{i}^{n}
\label{eratio} \end{eqnarray}
where the angle  $\psi_{i}^{n} = n\theta_j~-~i\varpi_1~+~(~i~-~2n~)~\varpi_2$ 

\noindent $\equiv~(~2n~-~i~)~\phi_3~-~(~n~- ~i~)~\phi_1.$

\noindent For the self-similar migration we consider,  $\phi_1$ is assumed to be very close to $\pi$ and $\phi_3$ very  close to $0.$
As the angles appearing in (\ref{eratio}) are linear combinations of these angles, the right hand side is readily evaluated in this limit.
The deviations of $\phi_1$ and $\phi_3$ from $\pi$ and $0$ are small, vanishing in the limit of large migration time
and changing no faster than on that time  scale. Accordingly in that limit,  $d \beta  /dt$ may be neglected in (\ref{eratio}),
which, as $\alpha$ is determined by the resonance condition,  reduces to  an algebraic equation for the eccentricity ratio $e_1/e_2.$  
For details about the  determination of the $F_{i}^{n}$  for the second order resonances we consider below, see Appendix \ref{MD}.

\subsection{Resonant angles}
Use of  (\ref{ejcons}) and (\ref{eratio}) enables the determination of $e_1$ and $e_2$ in terms of the migration and circularization
time scales.  When these are comparable for both  planets, the eccentricities will scale as the square root of the ratio of the 
characteristic circularisation time scale to
the characteristic  migration time scale (e.g. Papaloizou 2003). 
The deviations of $\phi_1$ and $\phi_3$ from $\pi$ and $0$ may then be determined from the equations for $de_i/dt$
obtained from (\ref{REM}).  In the limit of small eccentricities, for second order resonances
 these are found to  be on the order of  ${\rm M}_*/(m_i n_i \tau_{c,i})$  which can be small
for sufficiently large evolution times. The corresponding quantity  for first order resonances is 
 ${\rm M}_*/(m_i n_i \sqrt{ \tau_{mig,i} \tau_{c,i}})$ which is smaller  by a factor $\sim 1/e_i.$ Accordingly, solutions
 with small eccentricities are   obtainable  at more  rapid migration rates  in that case.

\subsection{Behaviour at small eccentricities}\label{Behaviour}
In  the case of first order resonances, resonant angles may continue to librate even when the period ratios 
differ  significantly from strict commensurability, being driven away for example by tidal effects causing orbital
circularization ( Papaloizou \& Terquem 2010, Papaloizou 2011), even when the magnitude of the eccentricities
approaches zero. However, because $R_D$ is of higher order in the eccentricities, this does not  occur
for higher order resonances and the period ratio has to remain close to strict  commensurability for the resonant angles to librate.
For second order resonances, one can estimate the possible deviation, in the limit of zero eccentricities,  by considering the terms 
proportional to either $m_1$ or $m_2$ in Equation (\ref{eratio}). When the resonant angles are constant, these terms are both
equal to  $(p+2) n_2 - p n_1.$  For the 5:3, 7:5  and 9:7  resonances  appearing in simulations below, for $m_1=m_2,$
we find   $n_1/n_2 - (p+2)/p \sim K(m_1/M_{\oplus})(M_{\odot}/M_{*}) , $ where $K =3\times 10^{-5} , 3.5\times 10^{-5}, $ and $4.2\times 10^{-5}$ 
for $p=3,5 $ and $7$ respectively. This indicates that  $n_1/n_2 - (p+2)/p  < \sim 1.5\times 10^{-4}$ in the Super-Earth regime.

However, departures from strict commensurability can increase at larger eccentricities as the magnitude of the 
resonant part of the disturbing function increases.   Then from Equation (8.58) of  \citet{Murray1999}, 
 we estimate  $n_1/n_2 - (p+2)/p \sim K_e( e_1/0.1)\sqrt{(m_1/M_{\oplus})  (M_{\odot}/M_{*})  },$
 where $K_e =1.55\times 10^{-3} , 2.1\times 10^{-3}, $ and $2.6\times 10^{-3}$
 for $p= 3, 5, $ and $7$ respectively. 
 This indicates a libration width  more than an order of magnitude larger for eccentricities $\sim 0.1.$

\subsection{Higher order resonances}\label{Higherorder}
Although the main focus of this section has been on second order resonances, we have observed the occurrence of 
even higher order resonances in a few of our simulations.  We comment that the discussion of second order resonances
can be generalised to apply to higher order resonances. For a $k$-th order resonance the angles $\phi_1$ and $\phi_3$
are modified to become 
\begin{eqnarray}
\hspace{-7mm}&&\phi_1 = (p+k) \lambda_2 - p \lambda_1 - k \varpi_1 {\hspace{3mm} \rm and}\label{eq:phi_k1} \\
\hspace{-7mm}&& \phi_3 = (p+k) \lambda_2 - p \lambda_1 - (k-1) \varpi_1 -\varpi_2 . \label{eq:phi_k3} 
\end{eqnarray}
The formalism  developed from  (\ref{REM}) and (\ref{RD}) can  be applied as before with all angles now expressed
as  linear combinations of the new angles.  However,  the form of the $F^n_i,$ for $n\ne 0$   
is at least of order  $k,$ whereas when $n =0$ corresponding to the secular terms, as in the case of second order resonances
there are second order terms. 
The above means that when the eccentricity is not too small, the resonant angle deviations 
are expected to be  $\sim [ M_*/(m_i n_i  \tau_{mig,i})]  (\tau_{mig,i}/  \tau_{c,i})^{k/2},$
whereas for small eccentricities the system is expected to be dominated by secular interaction.

\section{Numerical Simulations}\label{sec:sim_details}

We  have performed numerical simulations of a pair of coplanar  migrating low-mass planets using the N-body code REBOUND \citep{Rei2012}. 
Additional acceleration terms that model the effects of orbital migration and eccentricity damping  that can be assumed to  result from the interaction of the planets with a gaseous  protoplanetary disc \citep[see eg.][]{Lee2002, Snell2001} are included. 
The resulting   acceleration ${\bf \ddot{r}}$ of a planet is given by 
\begin{eqnarray}
{\bf \ddot{r}} = {\bf f_g} - \frac{ 2 {\bf {\dot r}}\cdot {\bf \hat{r}}}{\tau_c}+ \frac{ {\bf r} \times {\bf j}}{\tau_{mig} |{\bf r}|^2}=
{\bf f_g} - \frac{ 2 {\bf {\dot r}}\cdot {\bf \hat{r}}}{\tau_c} - \frac{{\bf  \dot {r}} \cdot {\bf \hat {\varphi}}}{\tau_{mig}}\ . 
\end{eqnarray}
where  $\tau_{mig}$ corresponds to the timescale for migration  adopted above and similarly $\tau_{c}$ to  the timescale for orbital  circularization.
Note that we have dropped the subscript, $i,$ that indicates the planet from these particular quantities.
The  force per mass arising from gravitational interactions with other bodies   is  ${\bf f_g},$ 
  ${\bf j}$ is the angular momentum  per unit  mass,  and ${\bf \hat {r}}$ and ${\bf {\hat{\varphi}}}$ are the unit vectors   in
  the radial and azimuthal directions for  polar coordinates with origin at the central star.

Depending on  the detailed properties of the disc, planets in  the mass range  we consider  can either  undergo \mbox{Type I}  migration in a linear regime, or  be in a nonlinear regime for which partial gap formation  and/or  wake-planet  interactions occur \citep{Baruteau2013}. 
Furthermore in the inner regions close to the star where truncation due to a magnetic field may occur,
the disc structure may be strongly affected and strong outward propagating density waves may be present.
Planetary  migration and circularization rates   depend   on  the 
radial form  of the state  variables and  the equation of state in the protoplanetary disc
 \citep [e.g.][]{Paardekooper2010}. These  time dependent quantities determine both the magnitude and sign of the migration rate
 which thus  have considerable uncertainty over the disc lifetime which is characteristically $10^{6-7}\ \rmn{yr}$.
 
 In order to perform simulations  on this characteristic  time scale,  we  have considered a very simple disc model
 for which $\tau_{mig,i}$ and $\tau_{c,i},$ $i=1,2,$ are constants exterior to some cavity radius with $1/\tau_{mig,i}$  and $1/\tau_{c,i}$ vanishing inside it in some  cases. In other cases circularization was allowed to continue
 for one or both planets while inside the cavity in order to 
  model  the situation  where that continues
 after migration ceases. This can  occur when the mean disc surface density  increases
 outwards \citep [e.g.][]{Paardekooper2010}.  
 We have then surveyed a  large range of possible values for $\tau_{mig,i}$ and $\tau_{c,i}/\tau_{mig,i}$.

\subsection{Numerical surveys }\label{Numsurveys}
Each of the surveys  that we present in this paper  includes  $20\times 20$ runs  each of which
is for a particular choice  of  $\tau_{mig,i}$   and  $\tau_{c,i}/\tau_{mig,i}.$
 These quantities are identified by the parameters $N_1$ and $N_2$ which are  generate values  for these  timescales 
 for a planet of mass $m_i$ through the equivalent quantities
\begin{eqnarray}
&&\tau_{mig}=1.62^{N_2} \times  10^3 \times \frac{M_\oplus}{m_i} \equiv \tau_{m0} \frac{M_\oplus}{m_i} \mathrm{yr} \hspace{3mm} {\rm and}\ \nonumber \\
&&\tau_{c}/\tau_{mig}=1.3^{N_1} \times 3\times 10^{-4}\ .
\label{migdef}
\end{eqnarray}
The values  for $N_1$ and $N_2$ adopted were zero or integers  in the interval $[0, 19].$ 
This leads to  a migration timescale in the range
 $(10^3  -  9.6\times 10^6) (M_{\oplus}/m_i)\ \mathrm{yr}$ and the ratio $\tau_{c}/\tau_{mig}$  
in the range $(3\times 10^{-4} - 4.4 \times 10^{-2})$.
 The ratio $\tau_{c}/\tau_{mig}$  is expected to be $\sim 2.5\times 10^{-3}$ for a disc with aspect ratio  $\sim 0.05$ from 
 a discussion of disc-planet interactions \citep[see e.g.][]{Pap2000}. This lies in the middle of the considered range for one earth mass.
We consider surveys for which the outer planet starts on a circular orbit at either $1\ \rmn{AU}$ or $5\ \rmn{AU}$ with the 
inner planet close to but outside either the 2:1 or the 3:2 resonance. In addition,  we  incorporate  an inner cavity  of radius $0.08\ \rmn{AU}$ inside which migration torques do not operate. 
But note that   our results can be scaled to other values using the  scaling of  the units of space and time available for 
systems governed by gravity (see Section \ref{initconfig}). 
Parameters associated with the surveys we carried out are listed in Table \ref{tab:runs}.

\begin{table}
 \begin{tabular}{|c|c|c|c|c|c|c|c|}
\hline
Survey label  &$ \frac{m_1}{M_{\oplus}}$ &$\frac{ m_2}{M_{\oplus}}$ &$\frac{a_1}{1{\rm AU}}$&$\frac{a_2}{1{\rm AU}}$&Extended &$ N_{sec}$ \\
\hline
1AU\_1\_4\_21 & 1&4& 0.6& 1& &15\\
\hline
1AU\_1\_4\_32 & 1&4&0.7&1&y & 10\\
\hline
1AU\_1\_2\_21 & 1&2&0.6&1&y &10\\
\hline
1AU\_1\_2\_32 & 1 &2&0.7&1& &2\\
\hline
1AU\_2\_1\_21 & 1&2&0.6&1& &  5\\
\hline
1AU\_2\_1\_32 & 1 &2&0.7&1& &3\\
\hline
5AU\_1\_4\_21 & 1&4& 3.0& 5& &6\\
\hline
5AU\_1\_4\_32 & 1&4&3.5&5& & 4\\
\hline
5AU\_1\_2\_21 & 1&2&3.0&5&&3\\
\hline
5AU\_1\_2\_32 & 1 &2&3.5&5& &6 \\
\hline
5AU\_2\_1\_21 & 1&2&3.0&5& &4\\
\hline
5AU\_2\_1\_32 & 1 &2&3.5 &5& &4\\
\hline

\end{tabular}
\caption{Parameters for the  surveys.
The first column gives the  survey label. Successive columns then give from left to right,   the masses of the inner and outer planets
in earth masses,
  the initial  semi-major axes of the inner and outer  planet  in AU,  and  y indicating  that the survey was  extended with
    one or both planets undergoing circularization inside the cavity.
    The final column gives  the number of second order resonances present at the end of all the runs constituting
the survey before any extension.}
\label{tab:runs}
\end{table}

For the first set of surveys,  when a planet  enters the inner cavity, the migration and circularization forces are switched off and it is only
 the gravitational forces  between  it and the  other planet and the central star which determines the subsequent evolution. 
 Runs are found to undergo little change in semi-major axis or period ratio once both planets are inside the cavity.   As these are the main quantities of interest,  the evolution can be  regarded as being 
 complete at this stage.  Accordingly we stopped a run once  both planets were interior to 0.075 AU.
A simulation time of  $10^7\ \rmn{yr}$ was found to be adequate to ensure the completion of all our runs.  

For the second  set of surveys, a survey of the type described above is extended from  the  point at which  both planets are in the cavity
with  circularization  forces  continuing   to apply  only for the outer planet but  with  migration  deactivated for both planets.
The  device of enabling circularization to continue for the outer planet  once convergent migration has ceased was adopted by
\citet{Baruteau2013}  to simulate the effects of wake-planet interactions. It  can  be regarded as equivalent to a situation
where there is a zero torque condition in the vicinity of  the cavity boundary with  circularization continuing  to operate there.
Such a cavity may be assumed to exist in practice as a consequence of a central magnetosphere or some other structural feature of the disc.
\begin{figure*}
 \centering
\includegraphics[width=8.75cm,height=10cm]{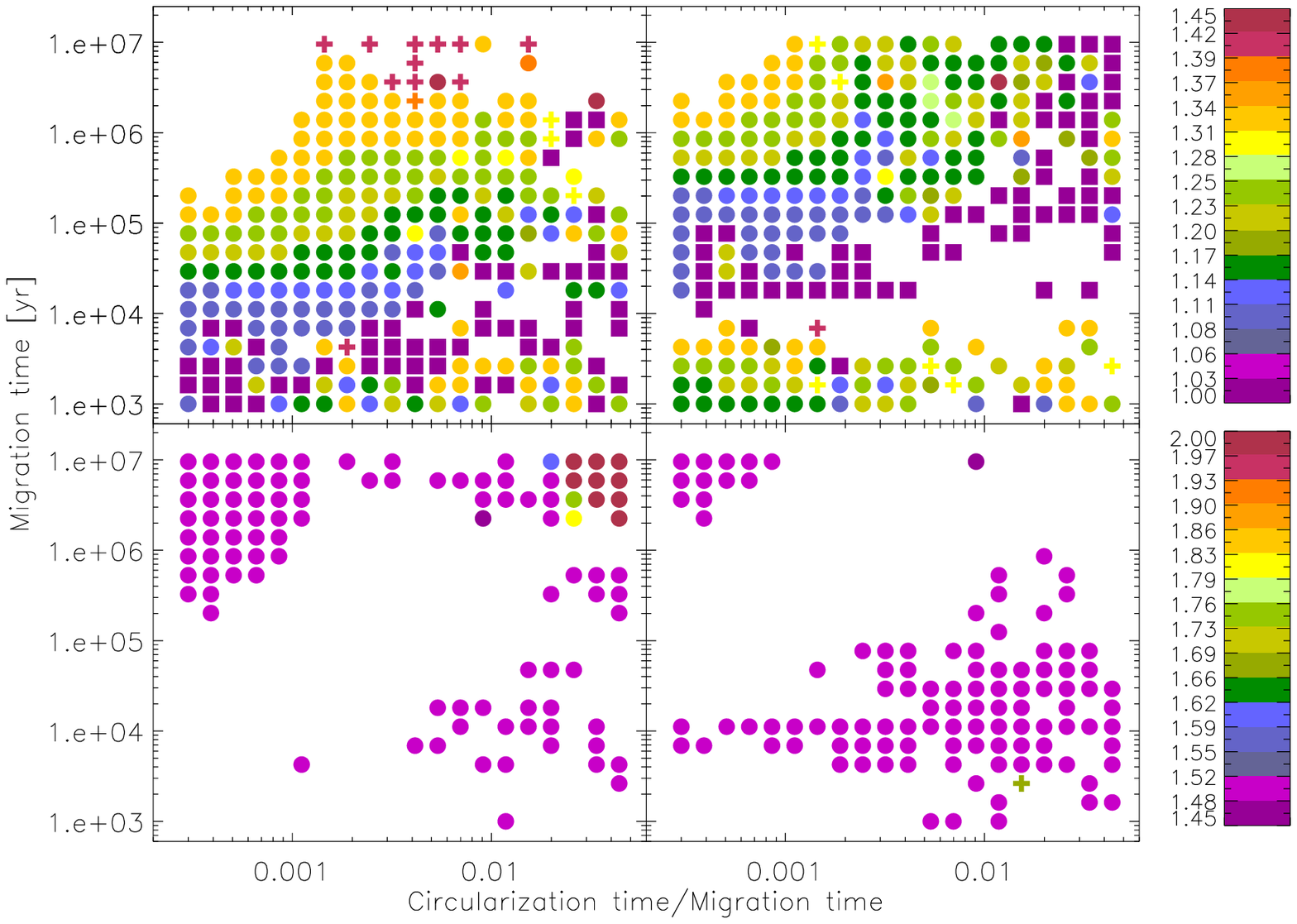}
\includegraphics[width=8.75cm, height=10cm]{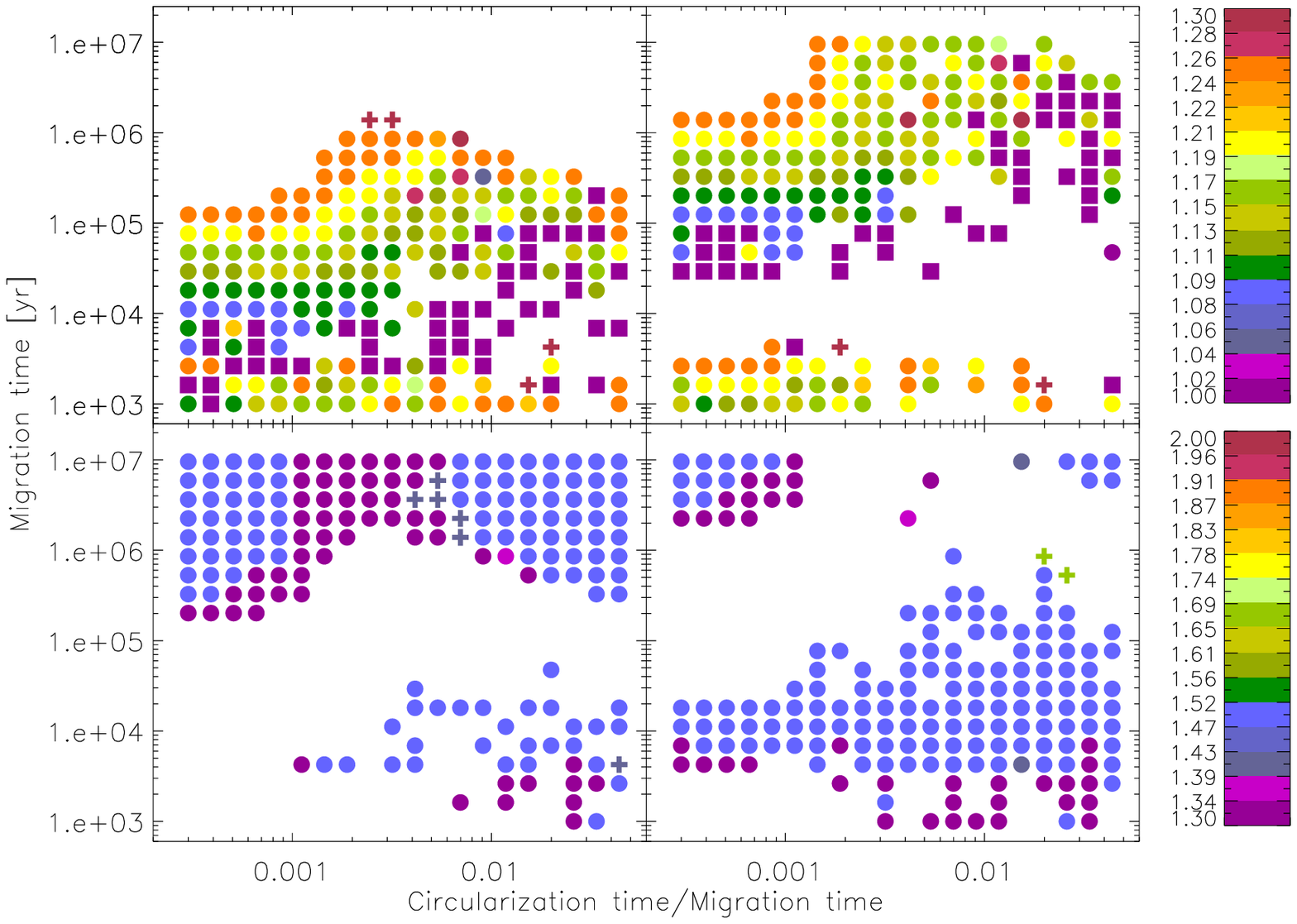}\\
\caption{Final period ratios for planet pairs initialised on orbits close to 2:1 (left  group of four  panels)
and 3:2 (right  group of four panels) with the inner planet mass being $1\ {\rm M}_\oplus$ and the outer mass being $4\ {\rm M}_\oplus$.
 For each group of four panels,  the upper and lower left panels  give results for the outer planet starting  at 1 AU and the upper and lower right
panels give results for the  outer planet starting  at 5 AU. These respectively correspond to the surveys $1{\rm AU}\_1\_4\_21,\hspace{2mm} 5{\rm AU}\_1\_4\_21,\hspace{2mm} 
1{\rm AU}\_1\_4\_32,$ and $\hspace{2mm}5{\rm AU}\_1\_4\_32.$
Period ratios are indicated in associated colour bars such that the larger values up to $2$ are indicated in the lower panels
and the smaller values down to $1$ are indicated in the upper panels.  Second order resonances are indicated by crosses (see text) 
 and the purple squares indicate collisions between the planets. Note that in this and other similar figures,
 the migration time plotted is $\tau_{m0}$ and the ratio of circularization time to migration time plotted 
is $\tau_c/\tau_{mig}$   which are both  independent of planet mass (see Equations (\ref{migdef})).
 For the simulations shown in the above panels the quantity,  $1/(1/\tau_{mig,2}-1/\tau_{mig,1}),$  which  estimates
 the initial convergence time for the outer and inner orbits is equal to  $ \tau_{m0}/3.$}
\label{fig:1_4_survey}
\end{figure*}

\subsection{Initial configuration}\label{initconfig}

We considered planets with masses of either 1, 2  or 4 earth masses ($\rmn{M}_{\oplus}$).  
The planets were initialised on circular coplanar orbits with the outer one being at 1 AU or 5 AU. 
The inner planet  is initialised on an orbit  which is  exterior to either  2:1 or 3:2  mean motion resonance.
The initial period ratio is $P_1/P_2=2.15$ in the former case and $P_1/P_2=1.7$  in the latter case.
Note that  the  system is invariant to multiplying all lengths by a factor $f_s$
and times, including $\tau_{mig,i}$ and $\tau_{c,i},$ $i=1,2,$  by a factor $(f_s)^{3/2}.$
These results can be extended to apply to cases with a different cavity radius.



\begin{figure*}
 \centering
\includegraphics[width=7cm]{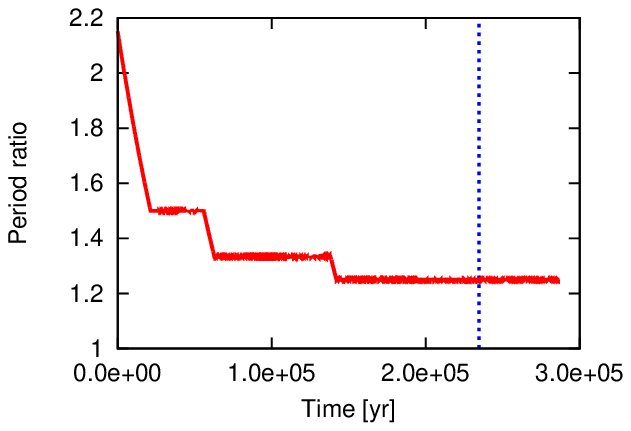}
\includegraphics[width=7cm]{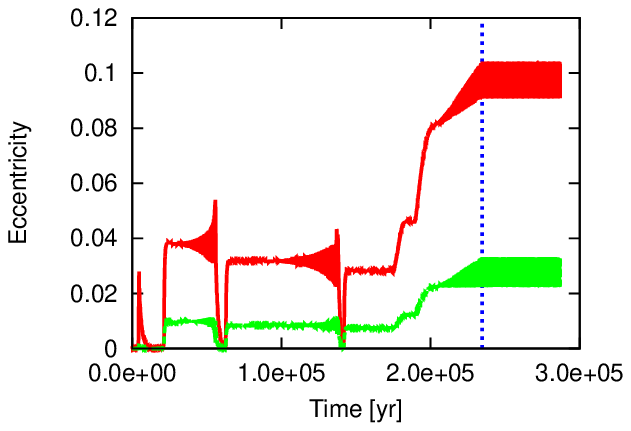} \\
\caption{The evolution of the period ratio (left panel) and eccentricity (right panel) for a run
contributing to the survey $1{\rm AU}\_1\_4\_21$ (see text).  The vertical line in this and similar figures indicates
the point at which both planets have entered the cavity.}
\label{fig:sequence_resonance}
\end{figure*}

\section{Results}\label{results}

\subsection{General features}\label{Genfeatures}

In all the unextended surveys listed in Table \ref{tab:runs}, the outcome is that either the planets are in resonance or they collide.
First and higher order resonances are possible but the latter are infrequent. We note that although planets
attaining first order resonances may undergo an instability that causes the resonance to be broken \citep{Gol2014}, another  resonance with smaller period ratio  is quickly attained. Thus most of the time is spent in some resonance unless the migration is fast enough to cause the planets to collide.

All our runs are stopped at that point. To illustrate a typical case, we describe the evolution of the period ratio and eccentricity for a run taken 
from the survey,  $1{\rm AU}\_1\_4\_21$, for which  the complete set of results are illustrated in Fig. \ref{fig:1_4_survey}.
The final period ratio is indicated in the eighth row and eighth column in the upper leftmost panel.
The results are illustrated in Fig. \ref{fig:sequence_resonance}. We recall that the outer planet mass was $4\ {\rm M}_{\oplus}$ and the inner planert mass $1\ {\rm M}_{\oplus}.$ 
Our migration prescription ensures convergent migration that first leads to trapping in the 3:2 resonance \citep[e.g.][]{Pap2005}.
 However, as the migration continues this is broken and a 4:3 resonance attained. That in turn is broken leading to a 5:4 resonance which remains until the pleanets
both enter the cavity. Note that towards and until just before the end of the run, the eccentricities increase.
 This is a common feature of the runs presented here, being due to the loss of the damping 
effect of the circularization of the inner planet while it is temporarily driven inwards by the outer planet.
Note that the transition from resonance to resonance seen here was also seen in simulations by \citet{Pap2009}.

\subsection{ Results of the surveys with   $(m_1, m_2)=(1, 4) \ {\rm M}_\oplus$ } \label{sec:1,4}
In Fig. \ref{fig:1_4_survey}  the
final period ratios for planet pairs initialised on orbits close to the 2:1  resonance are shown in the  left  set  of four  panels
and close to the 3:2 resonance in the right  set  of four panels. For each set of four the upper and lower left panels give the result
when the outer   planet starts at  1 AU and the upper and lower right
panels are  for the  outer planet starting  at 5 AU. Apart from a few exceptions associated with the planets scattering
as they enter the cavity, runs end with either the planets close to a resonance,  or with a collision.
Most of the resonances are first order and readily identified  in  Fig. \ref {fig:1_4_survey}  and similar figures below  with the 
help of the colour bars  (e.g.   for surveys starting close to the 2:1 commensurability illustrated 
in  Fig. \ref {fig:1_4_survey},   a  filled dark red circle  in the lower leftmost  panels corresponds to  2:1 resonance, 
a filled purple circle in those two panels  to the 3:2 resonance, a filled orange circle in the upper leftmost two panels to  4:3 resonance and a filled light green circle in the latter  panels to  5:4 resonance).

In general
planetary collisions are expected to occur for very fast migration (small migration time).
However, the rate  of migration required depends on the rate of orbital circularization,  decreasing  as the rate of circularization decreases. 
 This is because weak circularization  allows for the development of  highly  eccentric planetary orbits,  making  a collision  through orbit crossing more probable.
This trend  can be seen in Figure \ref{fig:1_4_survey} (and other surveys presented below). 
 As  migration times  increase,  collisions tend to  require  larger circularization times
in order to occur.  
 For example,  from the left group of four panels in Fig. \ref{fig:1_4_survey},  we can quantify the migration and circularization time limits for collision. 
In the left panels with the outer planet starting  at 1 AU,  collisions  can occur for  $\tau_c/\tau_{mig}\leq 0.01$
provided that  $\tau_{m0} \leq 10^4\ \rmn{yr}.$ This corresponds to the quantity,
 $1/(1/\tau_{mig,2}-1/\tau_{mig,1})= \tau_{m0}/3,$  which estimates the  convergence time  through migration of the orbits of the outer and inner planets 
(see equations (\ref {ejcons}) and (\ref{migdef})), being  $\leq 3\times 10^3\ \rmn{yr}.$     
  However,  when  $\tau_c/\tau_{mig} \sim  0.1$, collision can  occur for $\tau_{m0}\sim 10^6\ \rmn{yr}$. 
   Note that these times approximately   increase by a factor  $5^{3/2}$ for the right group of  panels with the outer planet starting at 5 AU.

To decide whether a run ended with a relatively rare second order commensurability, we adopted the same criterion as above that  
 fractional deviation  from commensurability  should be  $\ < 6\times 10^{-3}$. This was applied to all surveys.
 These were indicated in Fig. \ref {fig:1_4_survey}  and other similar figures by a cross. 
 Thus  for surveys with the outer planet starting close to the 2:1 resonance,  a  5:3 resonance  was indicated by an olive cross in the lower panel that is  second from the left, a  7:5
 resonance  by a burgundy cross in the two upper leftmost panels  and a 9:7 resonance by a yellow cross
 in the upper panel that is second from the left. These are discussed in more detail in Sections  \ref{Occurrence} and \ref{secondorderstudy} below.
  The survey $1{\rm AU}\_1\_4\_21$ had  also had runs that were associated with higher order resonances   with period ratios close to 
   10:7,  8:5  and 11:6.  These are discussed in more detail in Section \ref{Higherorderstudy}.

\subsection{Results of the surveys  with  $(m_1, m_2)=(1, 2)\ {\rm M}_\oplus$ and $ (m_1, m_2)=(2, 1)\ {\rm M}_\oplus$ } \label{sec:masses}

\begin{figure*}
 \centering
\includegraphics[width=8.75cm,height=10cm]{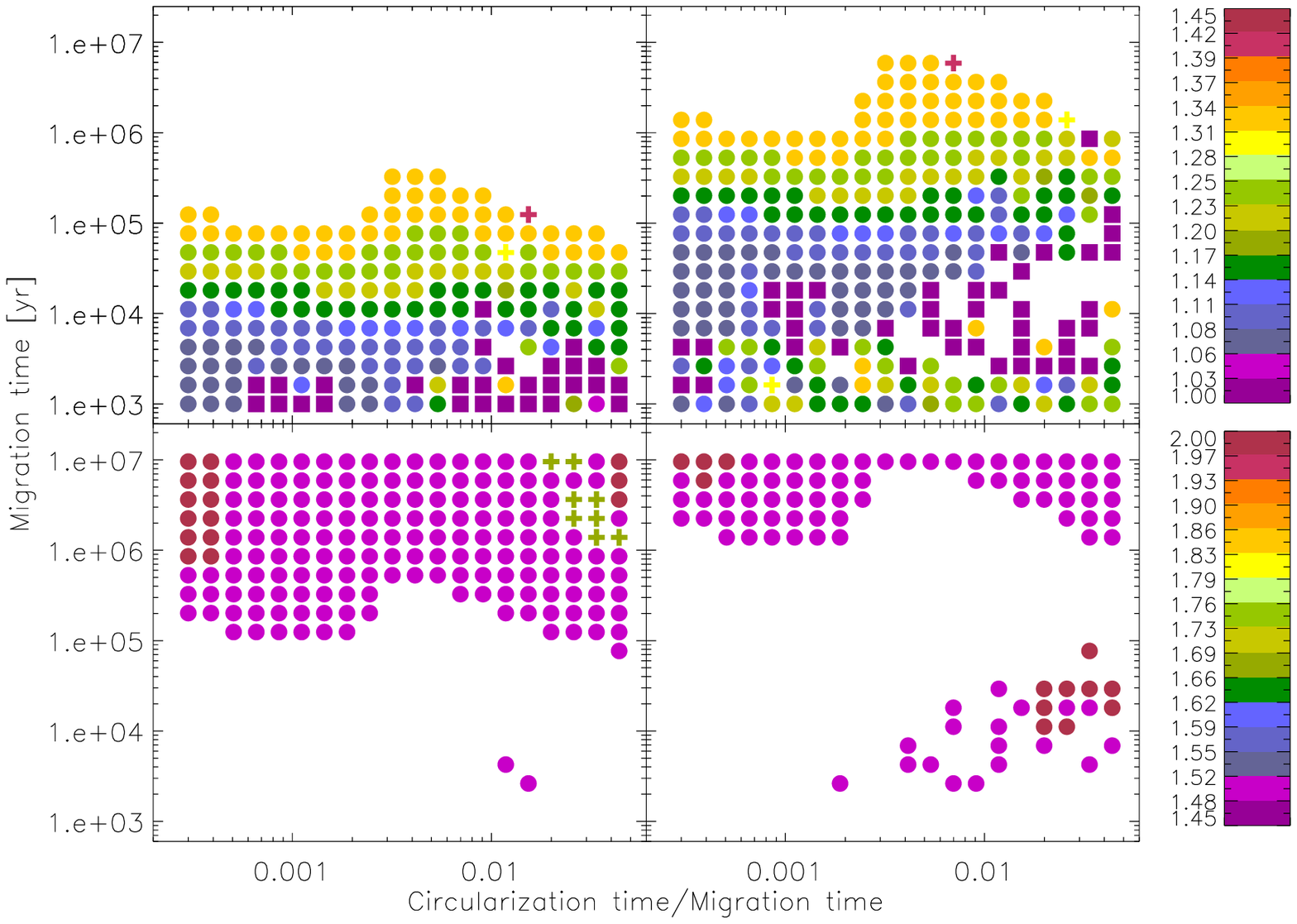} 
\includegraphics[width=8.75cm, height=10cm]{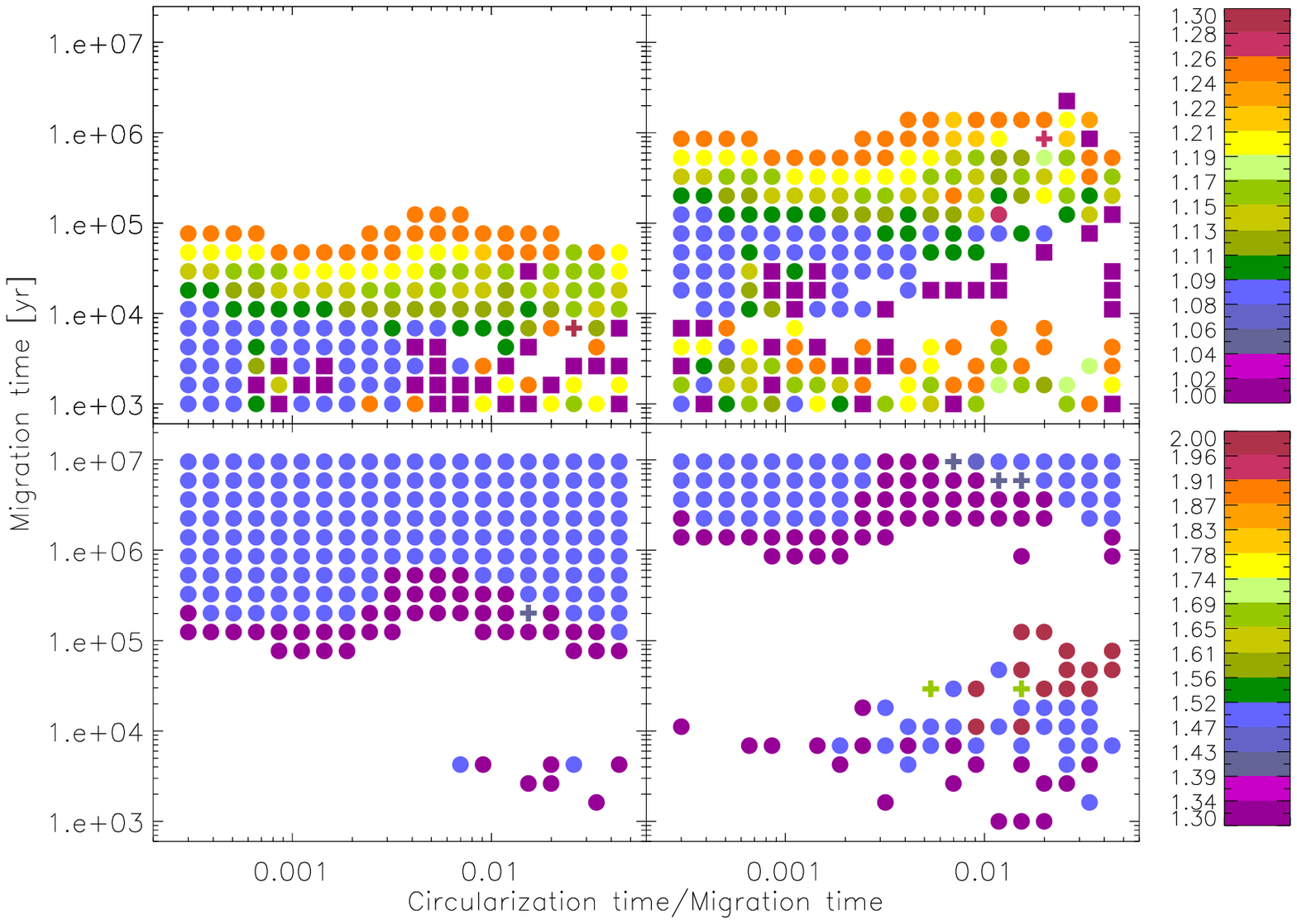}\\ 
\caption{
Final period ratios for planet pairs initialised on orbits close to 2:1 (left  group of four  panels) and 3:2 (right  group of four panels) with the inner planet mass being 1 ${\rm M}_\oplus$ and the outer mass being 2 ${\rm M}_\oplus$.  For each group of four panels,  the upper and lower left panels  give results for the outer planet starting  at 1 AU and the upper and lower right
panels give results for the  outer planet starting  at 5 AU. These respectively correspond to the surveys $1{\rm AU}\_1\_2\_21,\hspace{2mm} 5{\rm AU}\_1\_2\_21,\hspace{2mm} 
1{\rm AU}\_1\_2\_32,$ and 
$\hspace{2mm}5{\rm AU}\_1\_2\_32.$     
Period ratios are indicated in the associated colour bars such that the larger values up to $2$ are indicated in the lower panels
and the smaller values down to $1$ are indicated in the upper panels. 
 Second order resonances are indicated by crosses (see text)  and the purple squares indicate collisions between the planets. 
 For these cases, the quantity  $1/(1/\tau_{mig,2}-1/\tau_{mig,1}),$  which  estimates
 the initial convergence time for the outer and inner orbits is   $ \tau_{m0},$   which is the migration time plotted.
}
\label{fig:1_2_survey}
\end{figure*}

These surveys are carried out and the results are presented  in the same manner as the
surveys with $(m_1, m_2)=(1, 4)\ {\rm M}_\oplus,$
the only change being that they are carried out with different planet masses.
The qualitative description of the results is found to be the same.

 The results for the surveys with $(m_1,m_2)=(1, 2)\ {\rm M}_\oplus$
 are shown in Fig. \ref{fig:1_2_survey}. 
 \begin{figure*}
 \centering
\includegraphics[width=7cm]{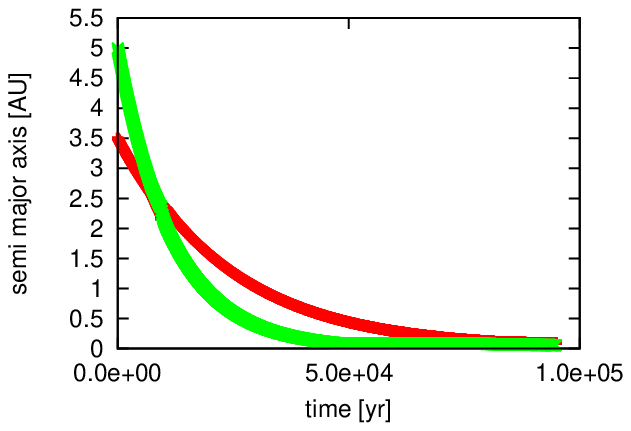} 
\includegraphics[width=7cm]{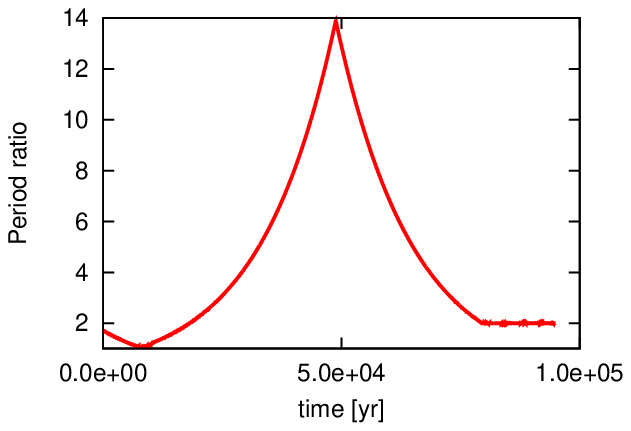} \\
\caption{The evolution of the period ratio (left panel) and eccentricity (right panel) for a run 
contributing to the survey $5{\rm AU}\_1\_2\_32 $(see text).  In this case the planets undergo orbit crossing and scatter
 causing  the inner and outer planets to be interchanged.  }
\label{fig:1_2_swap_example} 
\end{figure*}
\begin{figure*}
 \centering
\includegraphics[width=8.75cm,height=10cm]{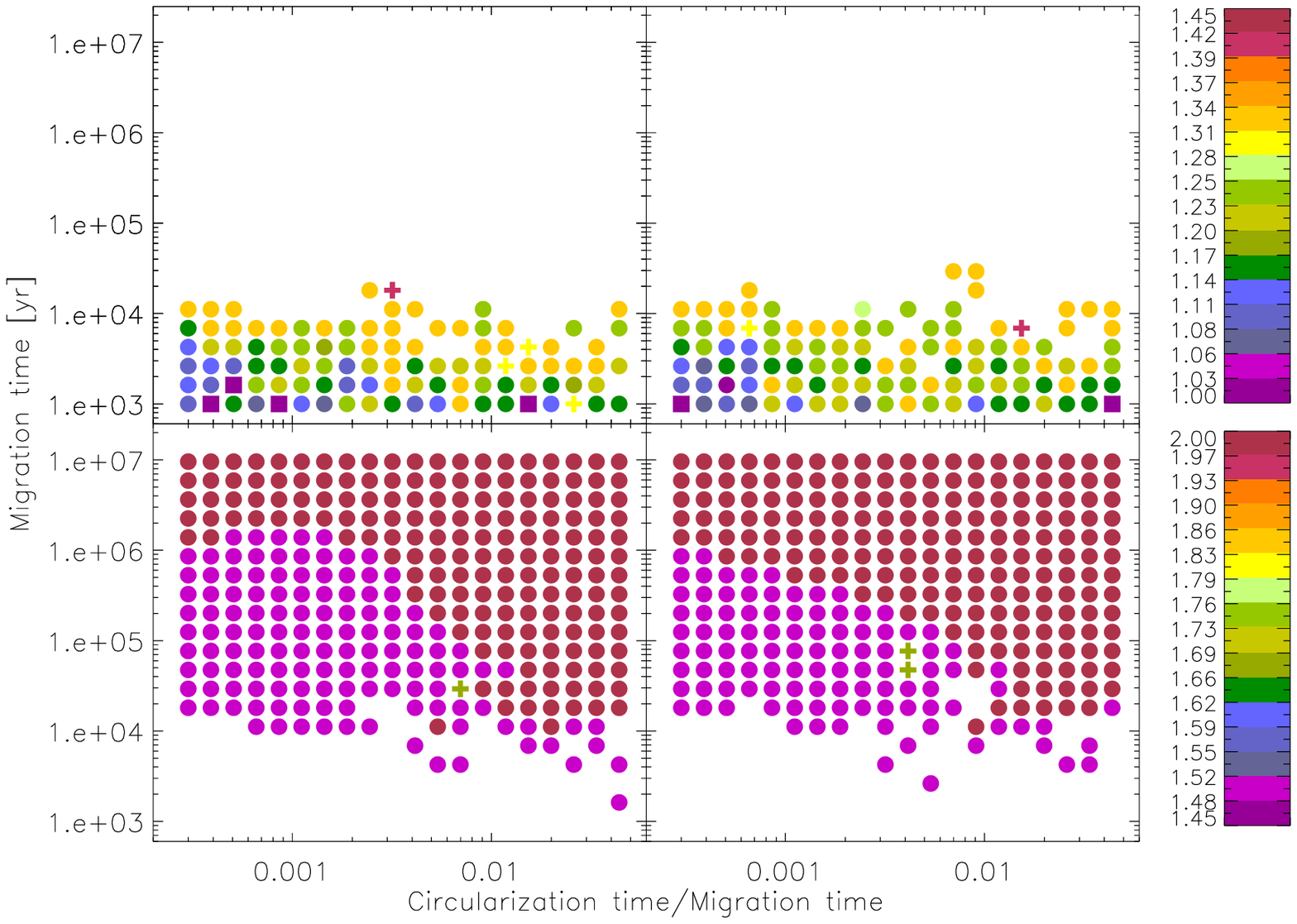} 
\includegraphics[width=8.75cm, height=10cm]{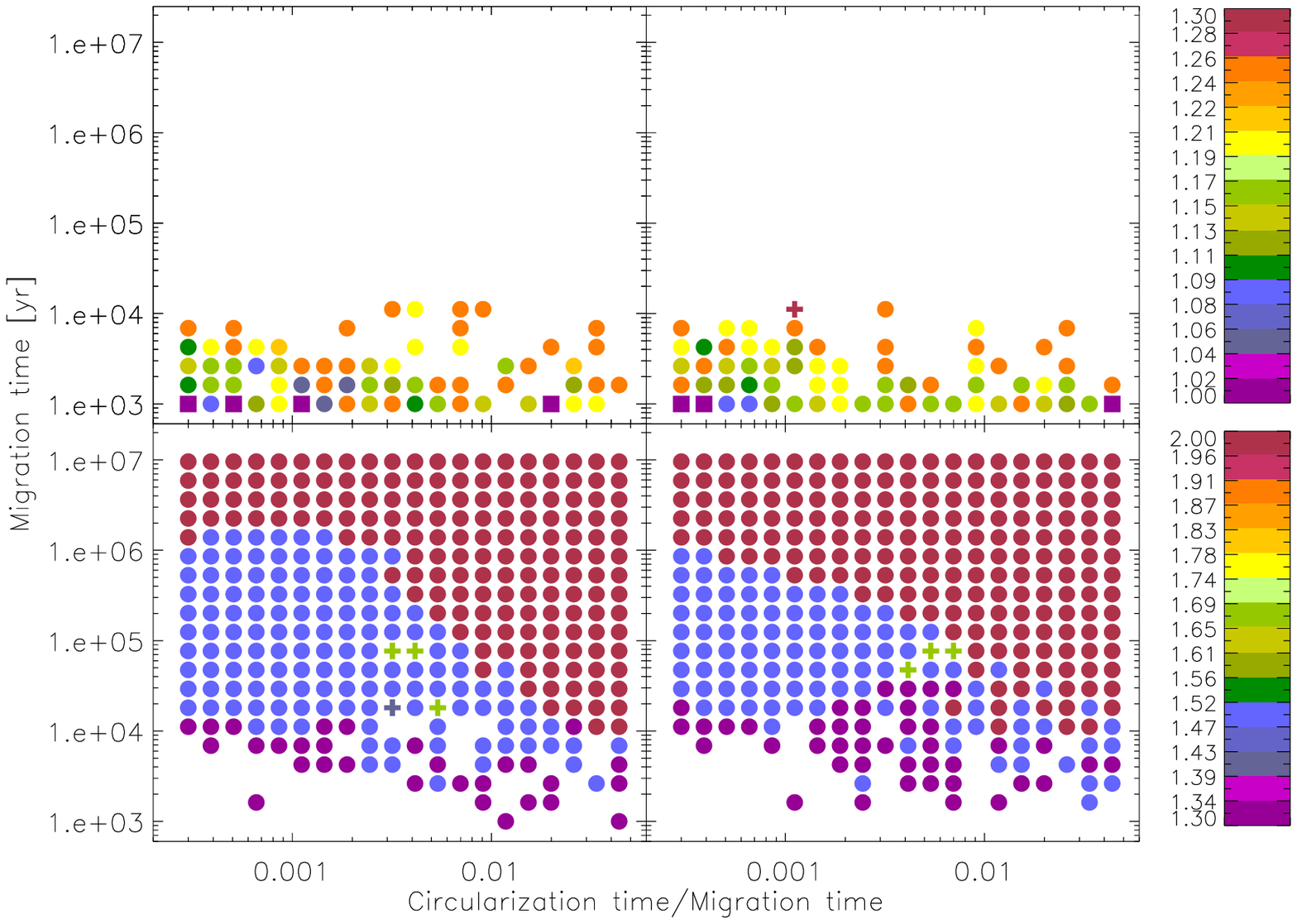}\\ 
\caption{
Final period ratios for planet pairs initialised on orbits close to 2:1 (left  group of four  panels) and 3:2 (right  group of four panels) with the inner planet mass being 2 ${\rm M}_\oplus$
and the outer mass being  1 ${\rm M}_\oplus$. For each group of four panels,  the upper and lower left panels  give results for the outer planet starting  at 1 AU and the upper and lower right
panels give results for the  outer planet starting  at 5 AU. These respectively correspond to the surveys $1{\rm AU}\_2\_1\_21,\hspace{2mm} 5{\rm AU}\_2\_1\_21,\hspace{2mm} 
1{\rm AU}\_2\_1\_32,$ and 
$\hspace{2mm}5{\rm AU}\_2\_1\_32.$     
Period ratios are indicated in associated colour bars such that the larger values up to $2$ are indicated in the lower panels
and the smaller values down to $1$ are indicated in the upper panels.  
Second order resonances are indicated by crosses (see text)  and the purple squares indicate collisions between the planets. 
The quantity  $1/(1/\tau_{mig,1}-1/\tau_{mig,2}),$  which for these cases estimates the initial
  divergence time for the outer and inner orbits is   $ \tau_{m0},$   which is the migration time plotted.
  }
\label{fig:2_1_survey}
\end{figure*}
These surveys produce a similar number of collisions, range of first order resonances  and second order resonances 
as the previous ones.  However, they differ in that a significant number of 2:1 commensurabilites are formed at high  migration rates and relatively
low circularization rates. These are seen in the second and fourth of the lower panels counted from the left of Fig. \ref {fig:1_2_survey}.
They are found to be associated with a scattering and interchange of the planets.
Figure \ref{fig:1_2_swap_example} shows the 
 evolution  for a run taken from  the survey $5{\rm AU}\_1\_2\_32.$  It is indicated in the ninth row from bottom and rightmost column of the rightmost lower panel of Fig. \ref{fig:1_2_survey}.
  In this run the planets undergo orbit crossing and scatter causing  the inner and outer planets to be interchanged and a large period ratio $\sim 14$ attained.  Subsequently convergent migration resumes and a 2:1 commensurability is formed.

Results for the surveys with   $(m_1,m_2)=(2, 1)\ \rmn{M}_\oplus$ are shown in Fig. \ref{fig:2_1_survey}.
In this case the general form of the orbital evolution differs from the previous cases.
As  the inner planet is more massive than the outer planet,  the initial inward migration is divergent.
 Only when the inner planet has reached the inner cavity while the outer planet is still migrating can 
 their semi-major axes start to approach each other and their period ratio starts to decrease so that a resonance can form.
 The nature of this evolution allows a significantly larger  number of 2:1 comensurabilities to result from systems with  larger migration rates,  as compared to the previous surveys.

Note that the simulations with the outer planet starting at 5 AU, would lead to the same results as in the case of the outer planet starting at 1 AU if the timescales were multiplied by a factor of $5^{3/2}$
and the inner cavity radius was also increased by a factor of five. However, because the inner cavity radius is in fact the same in both cases, the simulations with the outer planet starting at
5 AU map into simulations starting at 1 AU  with a reduced inner cavity radius. This results in a simulation that is  allowed to evolve for longer.
This has the consequence that
the planets end up closer together or with smaller period ratios than would occur if the simple scaling was assumed. This is apparent for example when the left and right panels in the leftmost group of four
 panels  in Fig. \ref{fig:2_1_survey} are compared.  However, the simple scaling is relatively  more closely adhered to when collisions are considered because the inner cavity plays
less of a role in those cases (see discussion of the results shown in  Figs \ref{fig:1_4_survey} in section \ref{sec:1,4} ).

\begin{figure*}
 \centering
\includegraphics[width=7cm]{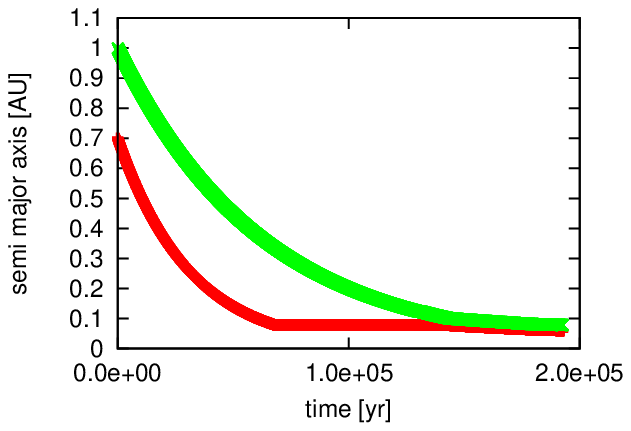} 
\includegraphics[width=7cm]{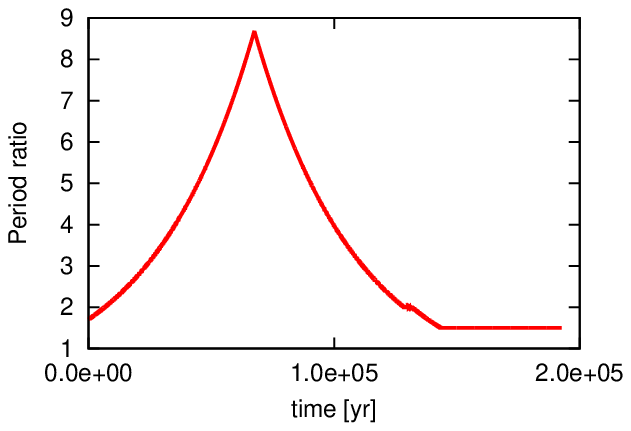} \\
\caption{ Evolution of the semi-major axis and the  period ratio for a run taken from the survey
$1{\rm AU}\_2\_1\_32$  for which the inner planet is  more massive (see text). } 
\label{fig:2_1_32_example}
\end{figure*}

In Figure \ref{fig:2_1_32_example}  we show the orbital evolution of a run  taken from the survey  $1{\rm AU}\_2\_1\_32.$ 
It corresponds to the run indicated in the eleventh row (from bottom) and eleventh column (from left) of the third lower panel from the left in  Fig. \ref{fig:2_1_survey}.
During the first part of the simulation, the inner planet migrates much faster inwards than the outer planet leading to a significant increase of their period ratio. This growth suddenly comes to a stop when the inner planet has reached the cavity and its migration  switches off.
 From this time on, the period ratio decreases  until a persisting  3:2  mean-motion resonance forms.  

In addition to producing more 2:1 commensurabilities significantly fewer collisions were found in these surveys as compared to the previously discussed ones.
However,  a comparable number of second order resonances were found.

\subsection{Occurence of second-order resonances}\label{Occurrence}

In Table \ref{tab:runs}
 we show the number of second-order resonances found in  all of  the  surveys discussed above. 
We detected  5:3, 7:5   and  9:7  resonances. These are indicated in the figures illustrating the final period ratios.

From  Table \ref{tab:runs}, there were $72$ occurrences out of a total of $4800$ simulations giving an incidence rate of $1.5\%.$
Although, these occurrences are rare, as they may contain useful information concerning their origin, we now examine some of them in more detail.

\subsubsection{Detailed study of second-order resonances}\label{secondorderstudy}

\begin{figure*}
 \centering
\includegraphics[width=4cm]{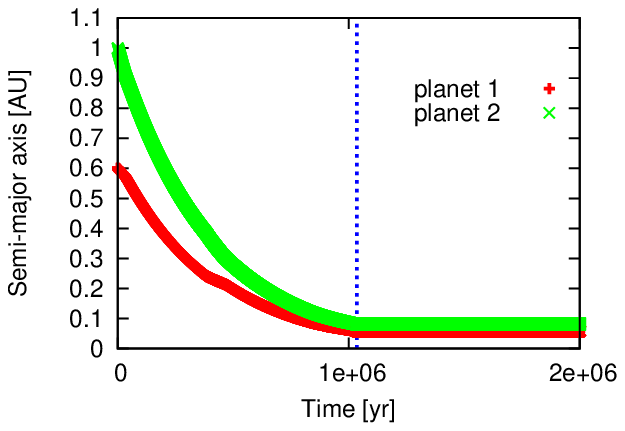}
\includegraphics[width=4cm]{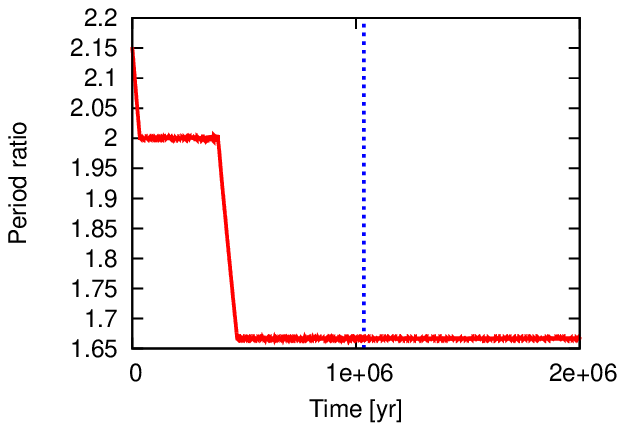}
\includegraphics[width=4cm]{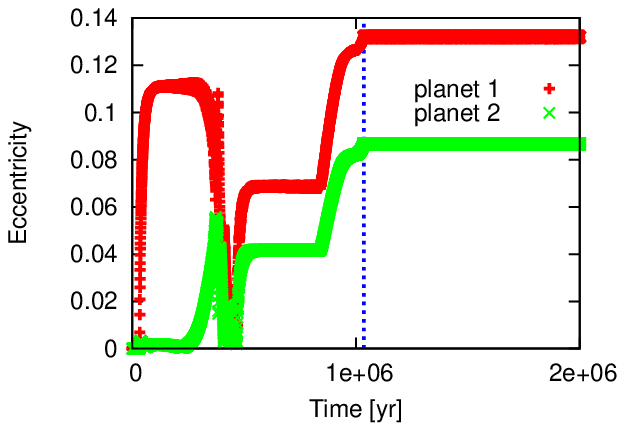}
\includegraphics[width=4cm]{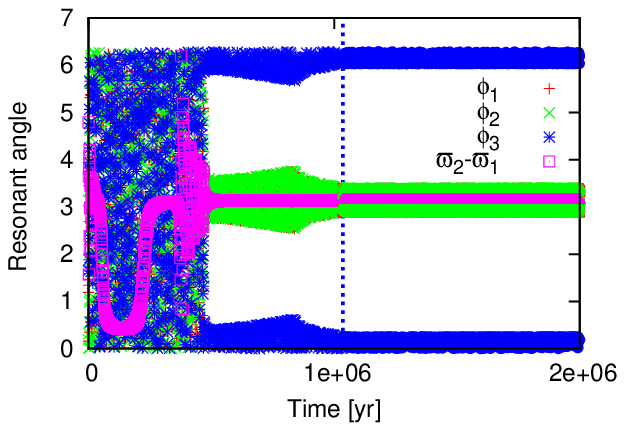}\\
\includegraphics[width=4cm]{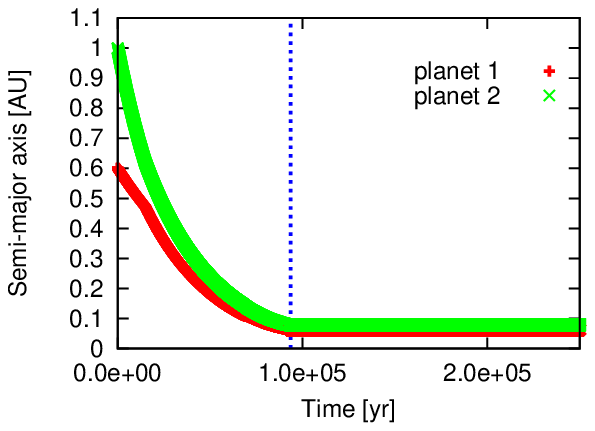}
\includegraphics[width=4cm]{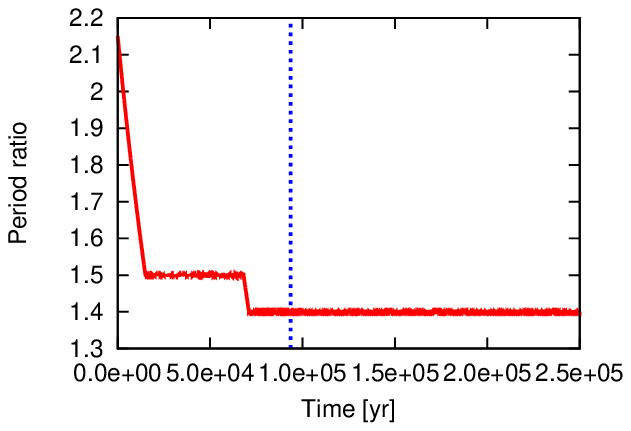}
\includegraphics[width=4cm]{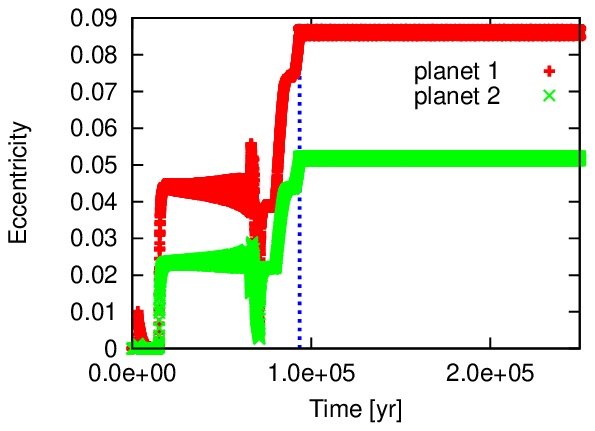}
\includegraphics[width=4cm]{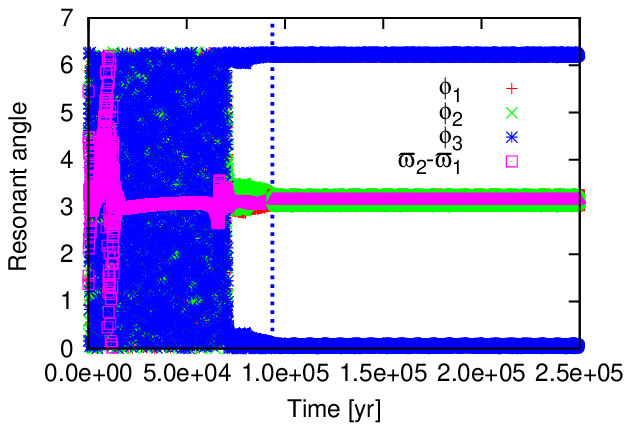}\\
\includegraphics[width=4cm]{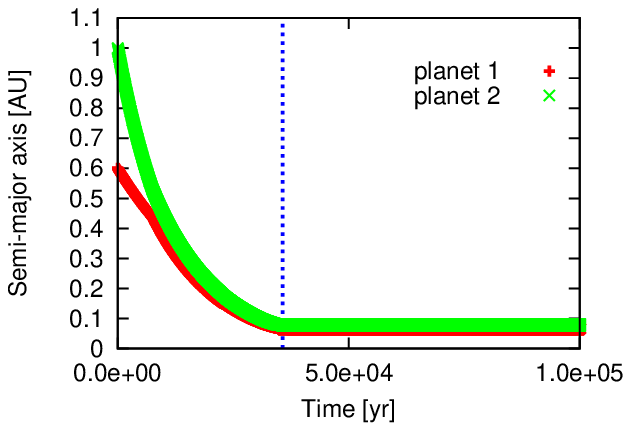}
\includegraphics[width=4cm]{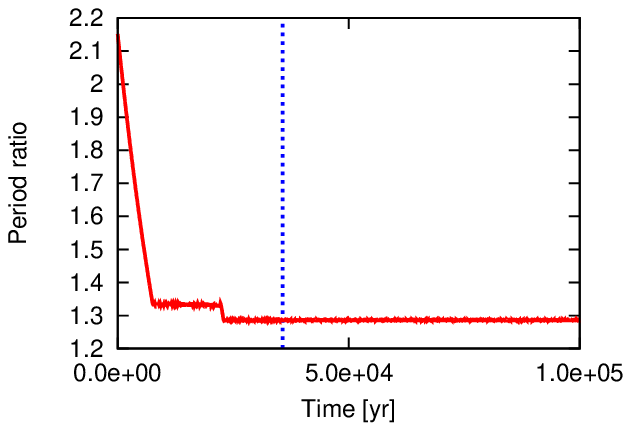}
\includegraphics[width=4cm]{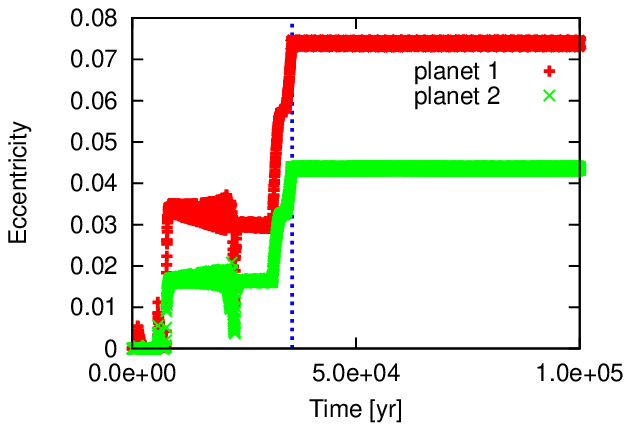}
\includegraphics[width=4cm]{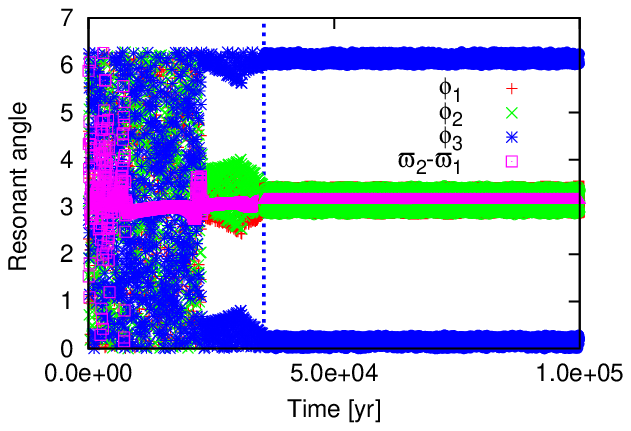}
\caption{Semi-major axis (first  column), period ratio (second  column), eccentricity (third column) and resonant angles $\phi_1$, $\phi_2$ and $\phi_3$ as well as $\varpi_2-\varpi_1$ (fourth column) for
runs that attained the
 commensurabilities  5:3   (upper panels),  7:5  (middle) and 9:7  (lower panels). The point at which both planets have entered  the inner
cavity  is indicated by the vertical  line. }
\label{fig:resonances_example}
\end{figure*}


\begin{table}
 \begin{tabular}{|l|l|l|l|}
\hline
Resonance & $e_1$&  $e_2/e_1$   & $r$   \\
\hline
5:3 &  0.13&  0.65&    0.571 \\
        &  0.07&  0.571&  0.557\\
\hline
7:5   &  0.087 &  0.575&  0.504\\
\hline
9:7 & 0.075  & 0.560&  0.456 \\
       &0.03      &0.5 &   0.481\\
\hline
\end{tabular}
\caption{Comparison of  numerical   (third column )  and  semi-analytically determined (fourth column) eccentricity ratios.
The resonance and value of the eccentricity of the inner planet are indicated in the first and second column, respectively.}
\label{tab:2nd_reso_comp}
\end{table}

We now examine some runs which ended with the planets in a  second order resonance in more detail.
For these the inner planet mass was 1 $\rmn{M}_\oplus$ and the outer planet mass was  2 ${\rm M}_\oplus$. The  outer planet started at 1 AU just outside the 2:1 resonance with the inner planet. They form part of the survey $1{\rm AU}\_1\_2\_21$ for which  results are illustrated in   the panels furthest to the left  of  Fig.
\ref{fig:1_2_survey}. We consider the unique runs in this survey that ended with the planets in 7:5 and 9:7 resonances
and the run corresponding to the second to last entry in the fifth row from the top of the lower panel furthest to the left.
 The latter (case A)  is one of several that ended in 5:3 resonance.
It had  $\tau_{mig}= 1.5\times 10^6(\rmn{M}_{\oplus}/m_i)\ {\rm yr}$ and $\tau_{c}/\tau_{mig}=0.03.$ For the 7:5 resonance case (case B), 
$\tau_{mig}= 1.2\times 10^5(\rmn{M}_{\oplus}/m_i))\ {\rm yr}$ and $\tau_{c}/\tau_{mig}=0.015$ and for the 9:7 case (case C)
$\tau_{mig}= 5\times 10^4(\rmn{M}_{\oplus}/m_i))\ {\rm yr}$ and $\tau_{c}/\tau_{mig}=0.012.$  

The evolution of the semi-major axes, period ratios,  eccentricities and resonant angles 
for these cases is illustrated in Fig. \ref{fig:resonances_example}.  At early times a first order resonance is attained, 2:1 in case A, 3:2 in case B and 4:3
in case C. This becomes unstable and the second order resonance is entered and maintained until the end of the run.
\noindent
$\phi_1$, $\phi_2$ and $\phi_3$ are the resonant angles as defined in Eqs. (\ref{eq:phi_1}) - (\ref{eq:phi_3}).
For all three simulations, when the second-order resonance occurs, $\phi_1$ and $\phi_2$ are around $\pi$ while $\phi_3$ is librating around 0 and $2\pi$. 
We recall that $\varpi_2-\varpi_1$ is the angle between the apsidal lines which is found to  librate around $\pi$. 

Averaging out short period oscillations, constant equilibrium eccentricities, as indicated in Section \ref{eccrat} are attained,  though this phase is short lived in case B.
The eccentricties  start to jump just after the inner planet starts to  enter the cavity.  This is because  when  it no longer intersects the disc,
migration and circularization then only  operate on the outer planet,  At later times, final constant values of the eccentricities are attained  that are larger than those applying when both planets were 
affected by migration and circularization.

We have tested the applicability of Equation (\ref{eratio} ) for the eccentricity ratio and Equation (\ref{ejcons}) which, with the help of Appendix \ref{MD}, enables the calculation
of the magnitues of the eccentricities during phases when the system undergoes self-similar migration, which we here identify with the phases when the eccentricities
are approximately constant with superposed small amplitude short period  oscillations. 
  Equation (\ref{ejcons}) then enables the calculation of the eccentricities as a function of the migration and circularization times.
In addition  comparison of the numerically determined eccentricity ratios with those obtained from the semi-analytic procedure is given in Table \ref{tab:2nd_reso_comp}. 
It is seen that there is good agreement, especially for the smaller  eccentricties. In these cases adequate  calculations 
can be performed taking only second order terms into account (see Appendix \ref{MD}). The absolute values of the eccentrities
are also found to agree with those implied by (\ref{ejcons}).


\begin{figure}
 \centering
\includegraphics[width=6cm]{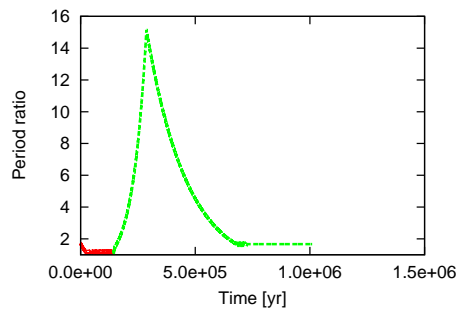}
\caption{The period ratio evolution  for the  run taken from $5{\rm AU}\_1\_4\_32$ that shows a scattering followed by the attainment of a 5:3  resonance (see text).}
\label{fig:resonances_53scattering}
\end{figure}

In Fig. \ref {fig:1_4_survey}    there are two  green  crosses corresponding to  5:3 resonances
 that can be seen in  in  the rightmost  lower  panel showing  the  survey with the  outer planet starting at 5 AU.
In these cases the relatively rapid  convergent migration rates result  in orbit crossing  leading to   the planets  changing  places.
The original inner planet is scattered outwards such that the system attains large period ratios $>14.$   
Convergent migration resumes after the now more massive inner planet enters the cavity resulting in the attainment of a 5:3 resonance.
The period ratio as a function of time for the case  with the shortest migration time is shown in Fig. \ref{fig:resonances_53scattering}.

\begin{figure*}
 \centering
\includegraphics[width=4cm]{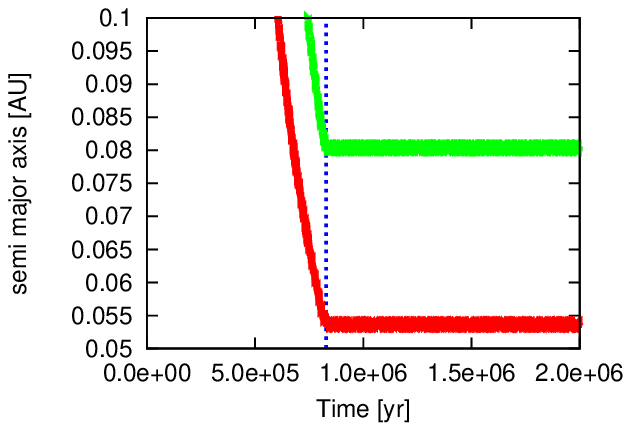}
\includegraphics[width=4cm]{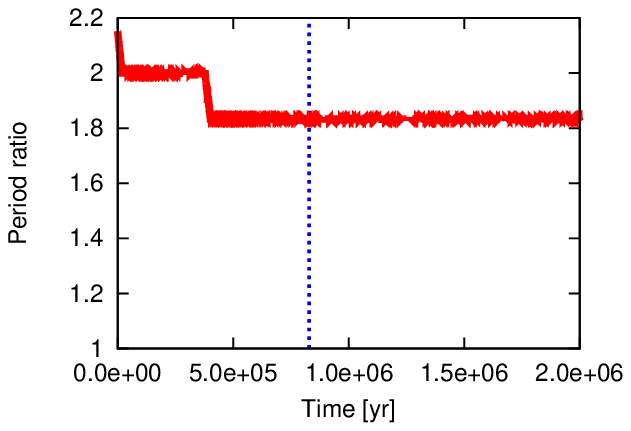}
\includegraphics[width=4cm]{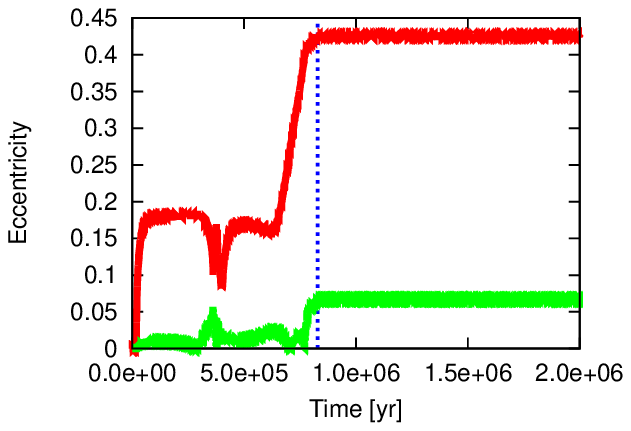}
\includegraphics[width=4cm]{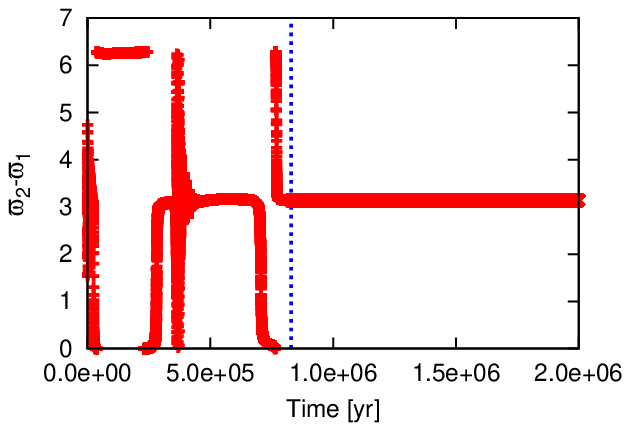}\\
\includegraphics[width=4cm]{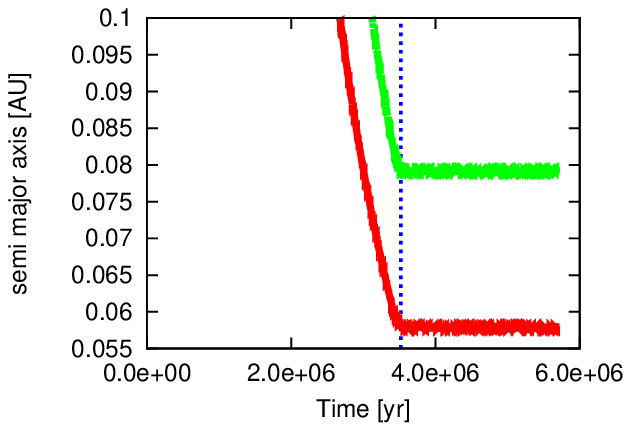}
\includegraphics[width=4cm]{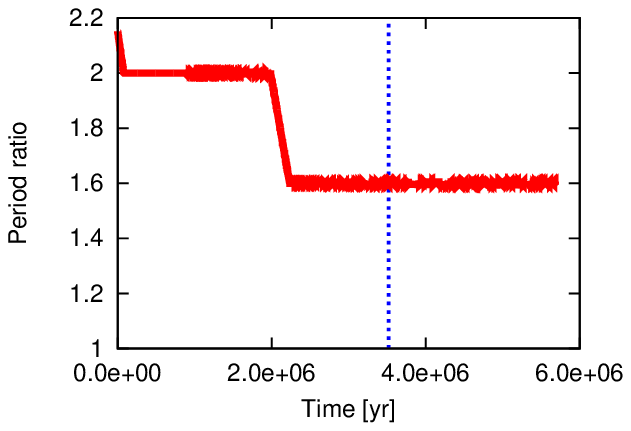}
\includegraphics[width=4cm]{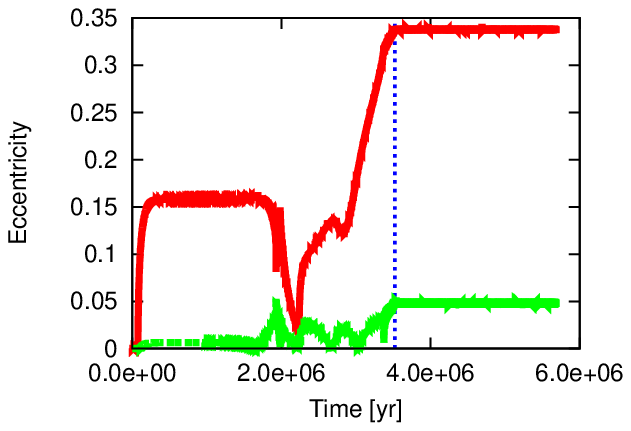}
\includegraphics[width=4cm]{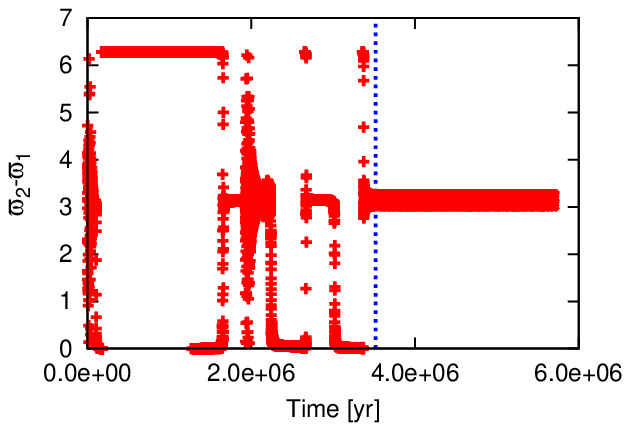}\\
\includegraphics[width=4cm]{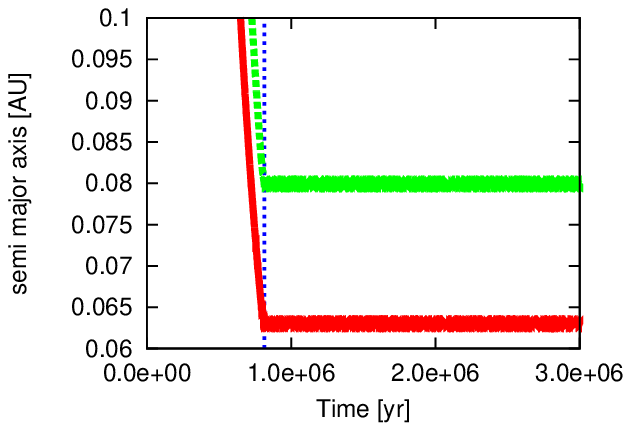}
\includegraphics[width=4cm]{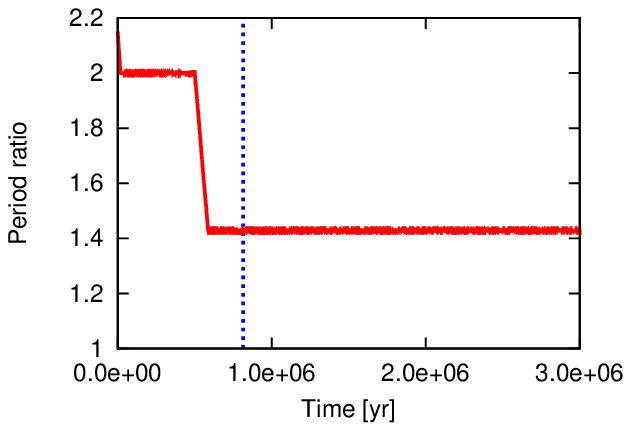}
\includegraphics[width=4cm]{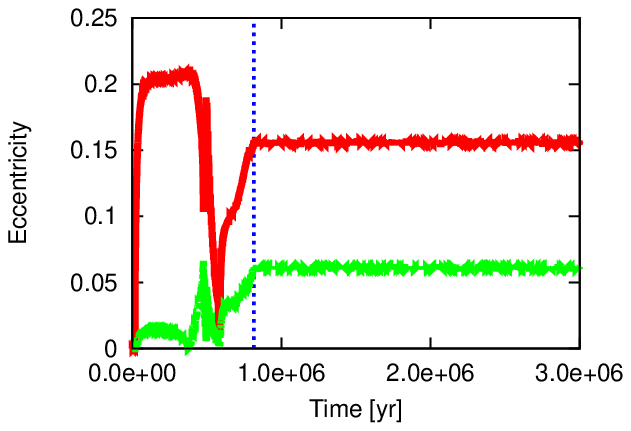}
\includegraphics[width=4cm]{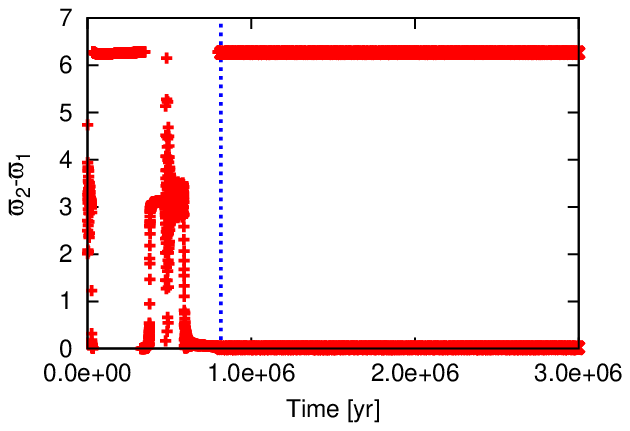}\\
\caption{Semi-major axis (first column), period ratio (second  column), eccentricity (third  column) and the angle  between the apsidal lines $\varpi_2 - \varphi_1$  (fourth  column) for
runs with  ultimate period ratios near  1.83 (upper panels),  1.6 (middle panels) and 1.427  (lower panels)  that  were carried out in the survey   $1{\rm AU}\_1\_4\_21.$}
\label{fig:resonances_examplea}
\end{figure*}

\subsection{Runs indicating higher order resonances}\label{Higherorderstudy}

\begin{figure*}
 \centering
\includegraphics[width=4cm]{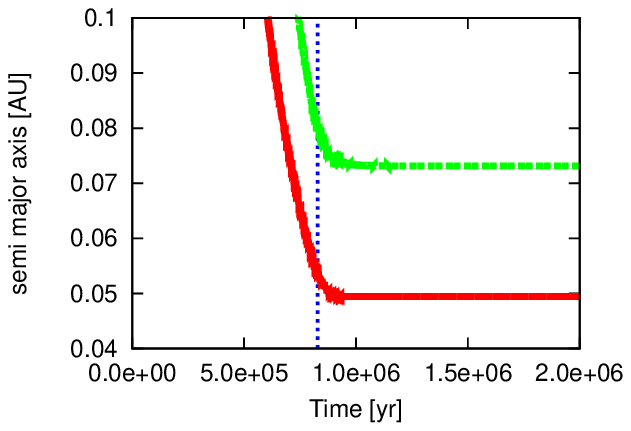}
\includegraphics[width=4cm]{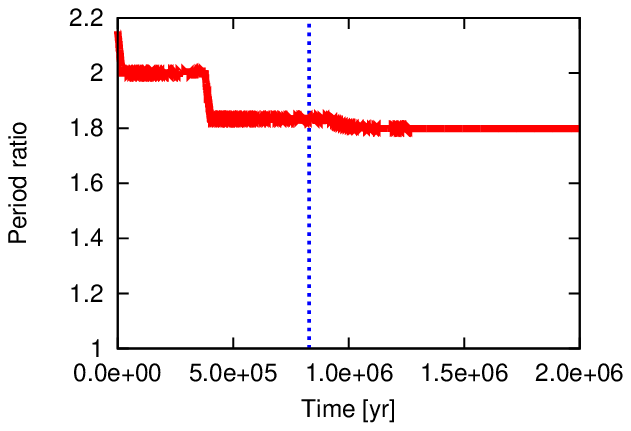}
\includegraphics[width=4cm]{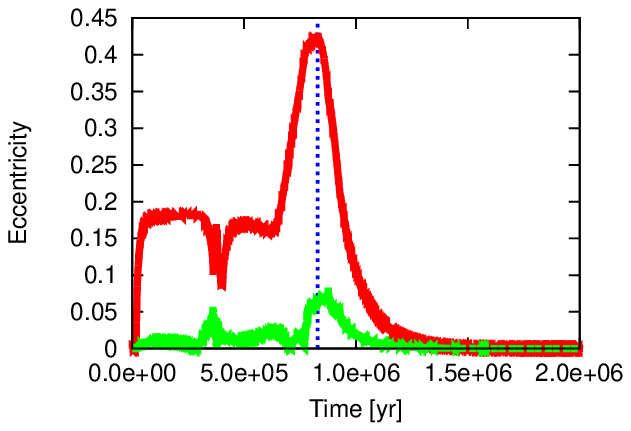}
\includegraphics[width=4cm]{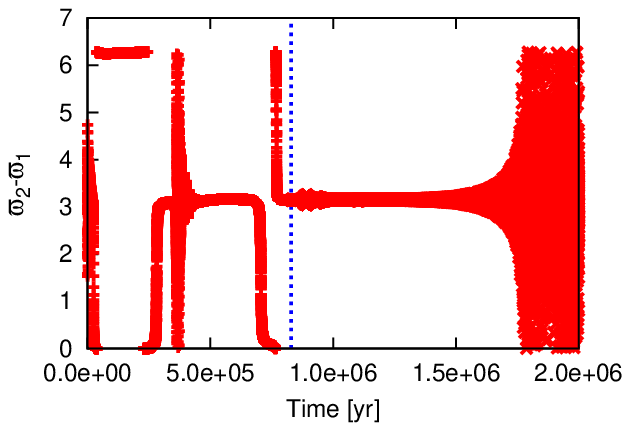}\\
\includegraphics[width=4cm]{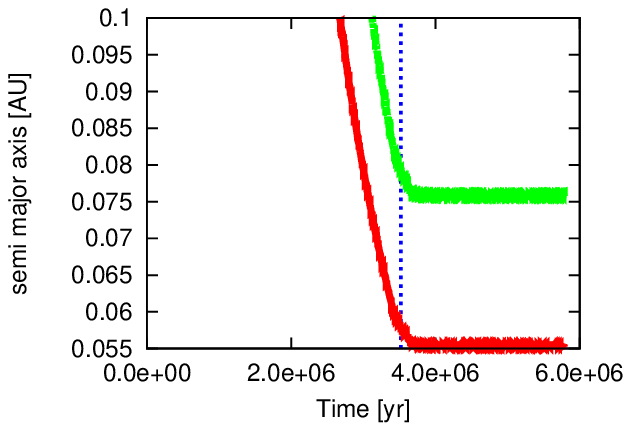}
\includegraphics[width=4cm]{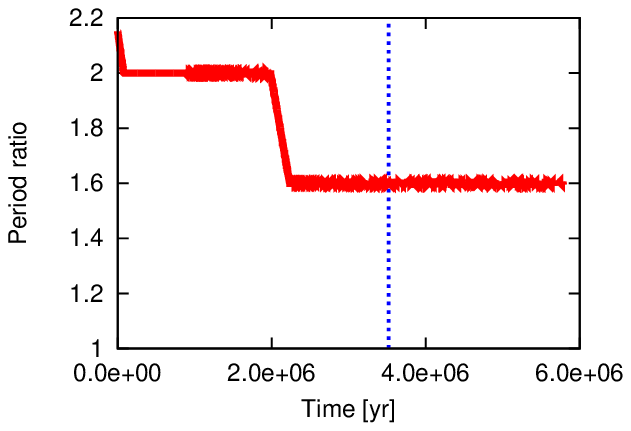}
\includegraphics[width=4cm]{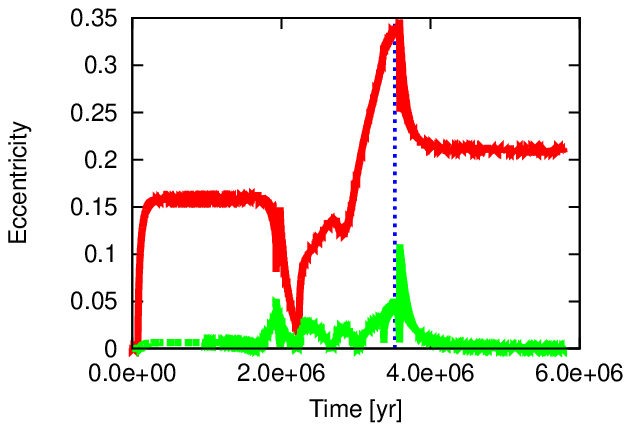}
\includegraphics[width=4cm]{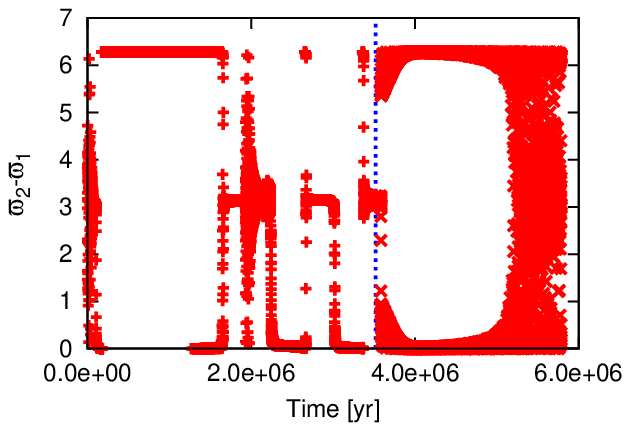}\\
\includegraphics[width=4cm]{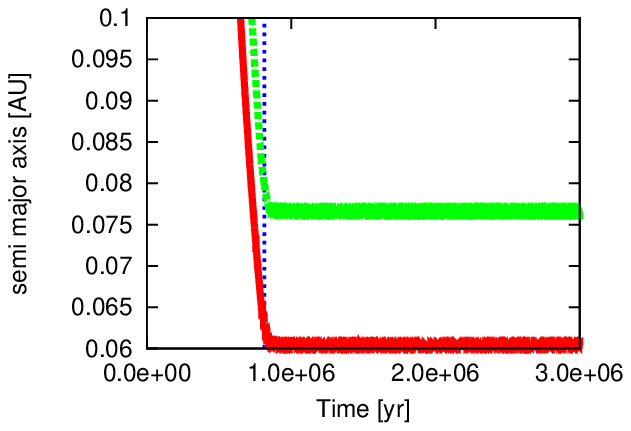}
\includegraphics[width=4cm]{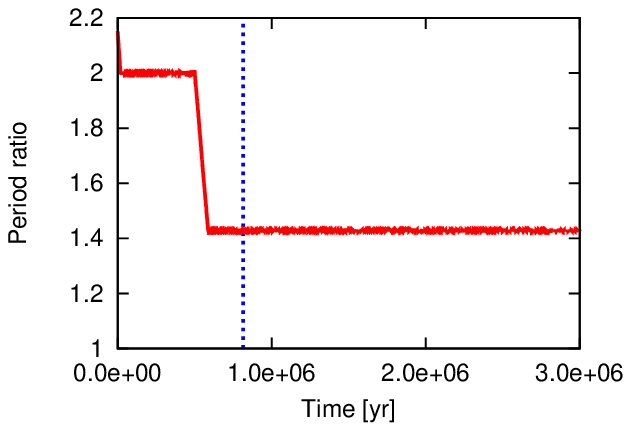}
\includegraphics[width=4cm]{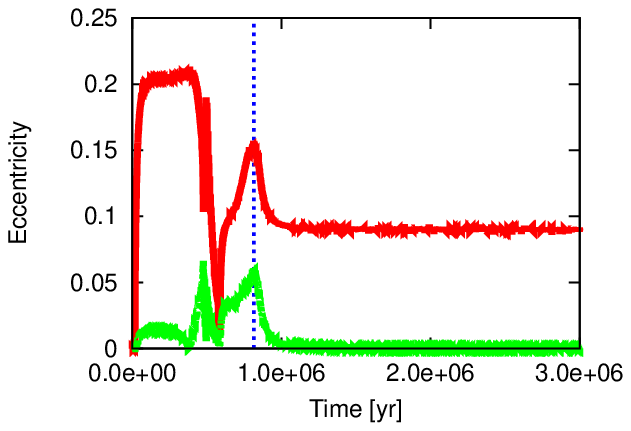}
\includegraphics[width=4cm]{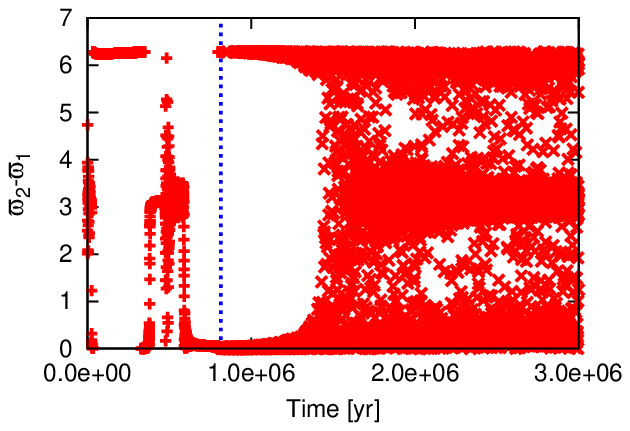}\\
\caption{As in Fig. \ref{fig:resonances_examplea} but  in this case after point where both planets enter the cavity, the runs are extended
with migration forces switched off  while  circularization forces continue to act only on the outer planet. }
\label{exampleeac}
\end{figure*}

To illustrate the potential presence of  higher order resonances  we illustrate the evolution of   runs taken from  the survey
$1{\rm AU}\_1\_4\_21.$  These were cases with final period  ratios 1.83 (case D), 1.6 (case E) and  1.427 (case F).
Case D  corresponds to the run shown in  the fourth row (from top) and third column from the end  in the lower leftmost  panel of Fig. \ref{fig:1_4_survey}.
Case  E corresponds to the run shown in  the top row  and fourth  column  from the end  in the lower leftmost  panel of Fig. \ref{fig:1_4_survey}.
Case  F  corresponds to the run shown in  the fourth row  and  second column from   the end in the upper leftmost  panel of Fig. \ref{fig:1_4_survey}.
The time independent evolution of the semi-major axes, eccentricities and angle between the apsidal lines for these cases is illustrated in Fig. \ref {fig:resonances_examplea}. 
Each of these runs initially attains a 2:1 resonance which is not sustained \citep{Gol2014}. The systems leave it and attain period ratios close to 11:6, 8:5 and 10:7 
in cases D, E and F respectively. The angle between the apsidal lines oscillates between $0$ and $\pi$ before settling on $\pi$ in cases D and E and $0$ in case F
after the planets have entered the cavity and forces arising from  migration and circularization were removed. 
Note that the eccentricities are large in these runs attaining $\sim 0.45$ in case D.

\begin{figure*}
 \centering
\includegraphics[width=8.75cm,height=10cm]{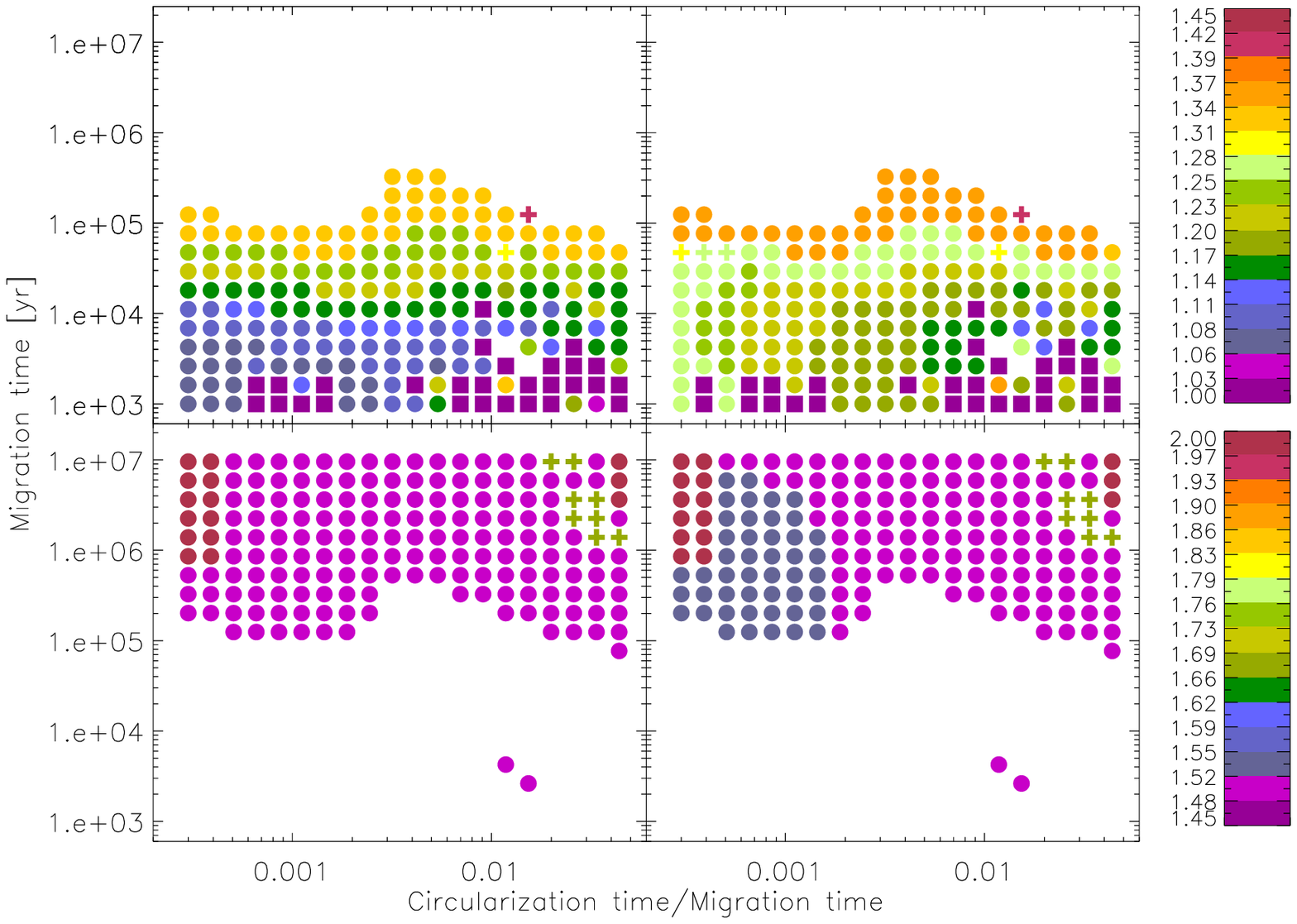}
\includegraphics[width=8.75cm,height=10cm]{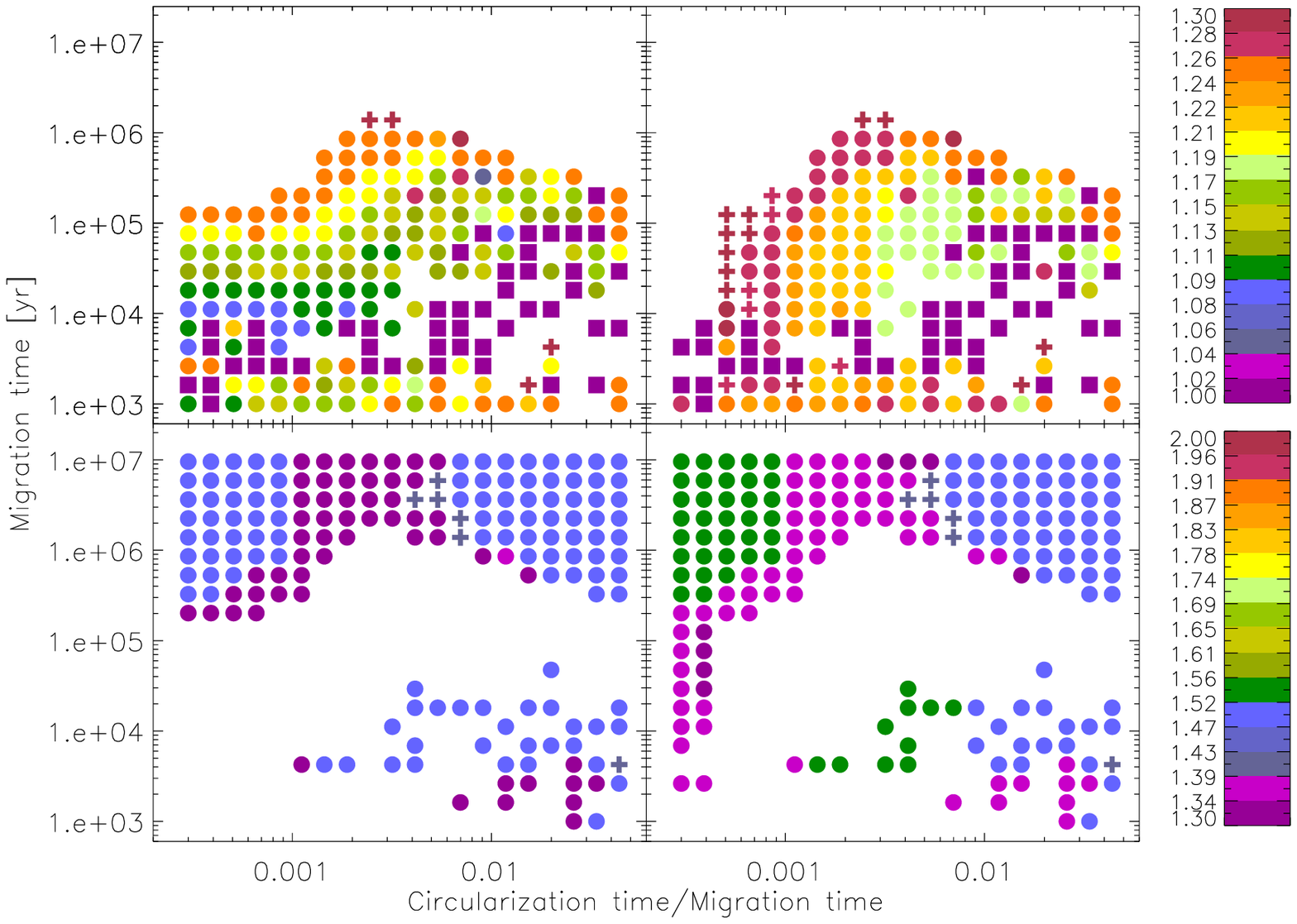} 
\caption{Final period ratios after extended runs
 for planet pairs initialised on orbits close to  2:1 (left hand set of four panels) and 3:2 
(right hand set of four  panels) are shown in the upper and lower right hand panels
of each group of four. The upper and lower left hand panels for each set show 
the period ratios at the commencement of the extension of the runs ( the result of the unextended  surveys
 $1{\rm AU}\_1\_2\_21$ and  $1{\rm AU}\_1\_4\_32$)  for comparison.
The inner planet mass is 1 $\rmn{M}_\oplus$ and the outer mass  2 $\rmn{M}_\oplus$ in the first case  and 4 $\rmn{M}_\oplus$ in the second.
 The outer planet starts at 1 AU   for all runs. 
 The purple squares again show simulations with collisions between the planets. 
}
\label{fig:1_4_longertime}
\end{figure*}

To  investigate the subsequent evolution of the period  ratios
that could occur if additional dissipative forces were imposed
we performed an additional set of runs for which circularization was allowed to continue to operate on the outer planet after the cavity was entered,
The evolution of these runs is illustrated in Fig. \ref{exampleeac}. It will be seen that the eccentricities decrease while changes to the period ratio are on average very small.
 The inner planet has the larger eccentricity while dissipation occurs only  for the outer planet.
The eccentricity of the inner planet decreases through secular  transfer  to the outer planet which  is indicated by maintenance of the angle between the apsidal lines at $0$ or $\pi.$
In case D the latter fails once both eccentricities become very small and the angle loses its significance.
In all cases the evolution  ceases  as the eccentricity of the outer planet approaches  zero.
During the later stages of evolution,  energy is removed from the orbit of the outer planet,  but not that of the inner planet,  causing  the period ratio to decrease very slightly.



\subsection{ Continuation of runs with orbital circularization but with no migration torques}\label{Continuationruns}

In this section, we describe  some  surveys of the second type that were continued after the planets have reached the inner cavity.
We begin by considering  extensions of the surveys  $1{\rm AU}\_1\_2\_21$ and
$1{\rm AU}\_1\_4\_32$  the results of which are shown in Figs \ref{fig:1_2_survey} and \ref{fig:1_4_survey}.

For the extensions to the runs, as described above  we  allow circularization  to continue only for the outer planet
  while migration is deactivated for both planets interior to the cavity. Each  run was then continued
for  a time which is the shorter of either $5 \times t_{max}$ or $2\times 10^7\ \rmn{yr},$ with $t_{max}$
being the duration of the unextended run.

Figure \ref{fig:1_4_longertime} shows the final period ratio at start of this continuation on the left panel and the final period ratio
 after continuing the simulations. By making a comparison  one can see  that the period ratios are increased. This is expected
because  during this evolution, the planetary systems have energy removed at constant angular momentum, and the theoretical 
expectation is that they should separate with period ratio increasing as a function of time \citep[e.g.][]{Pap2011}.

\begin{figure}
 \centering
\includegraphics[width=8.75cm,height=10cm]{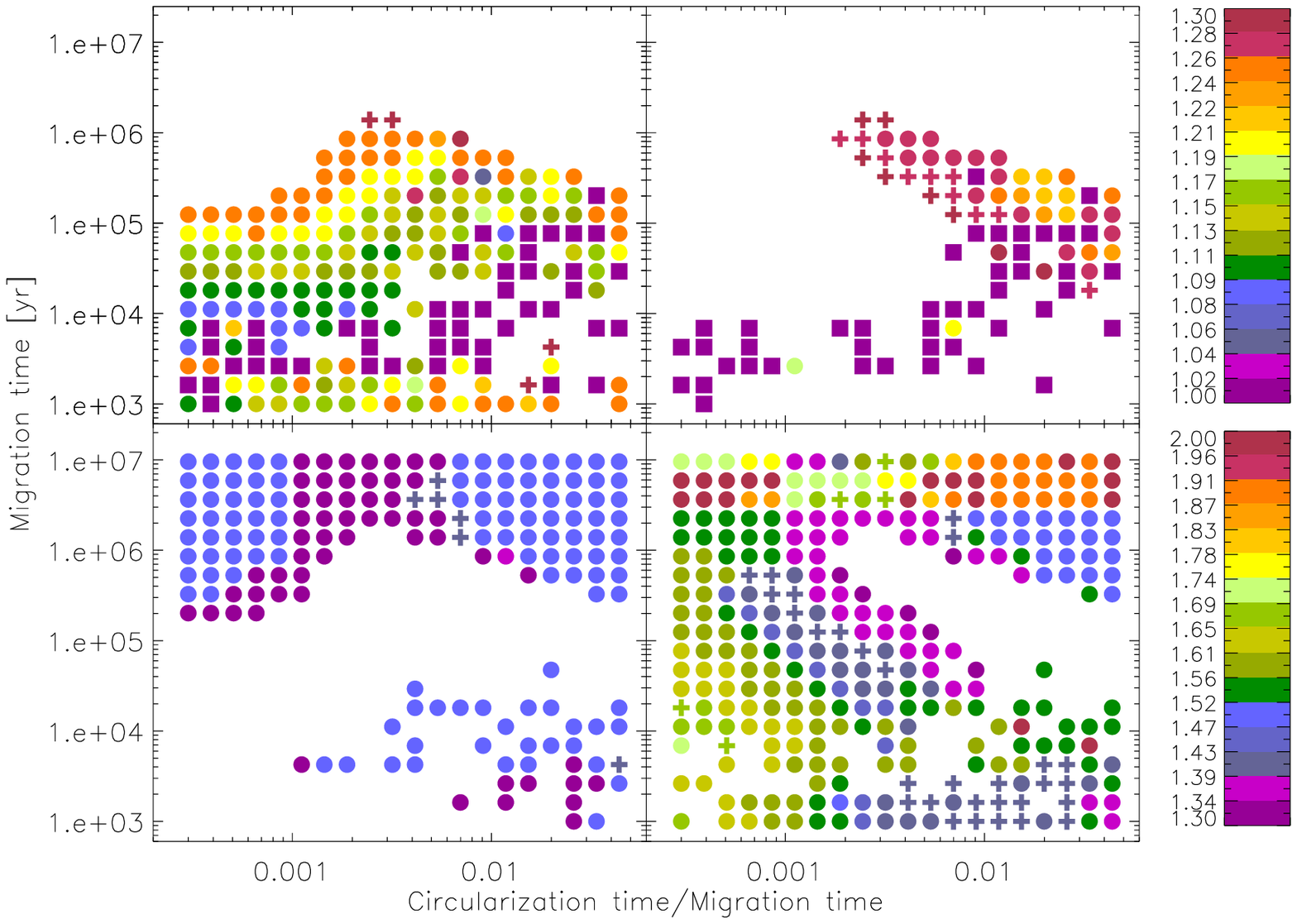}
\caption{Final period ratios after the  runs in the  survey  extended from $1{\rm AU}\_1\_4\_32$ 
that are  shown in Fig. \ref{fig:1_4_longertime} were further  extended for an additional time,  $0.1 t_{max},$
with a circularization time of $20$ orbits for both  planets,  are shown
 in the right hand panels.
 The upper and lower left hand panels  show
the period ratios at the commencement of the first extension of the runs (the result of the unextended  survey
  $1{\rm AU}\_1\_4\_32$).
 The purple squares show simulations with collisions between the planets.}
\label{fig:1_4_longertimeF}
\end{figure}

This is especially the case  for  small values of the ratio of the  circularization time to the migration time
 (see the  left hand sides of the panels).
 For example inspection
 of the two lower leftmost  panels of  Fig. \ref{fig:1_4_longertime} shows systems that were in 3:2 resonance increasing their period ratios in to the
 1.52 - 1.56 domain for both surveys illustrated. However, changes are much smaller for longer circularization times.   
In addition the period ratio range $1.7 - 1.8$ is empty. This can be rectified if shorter circularization times are used so inducing
a more rapid separation of the planetary systems. We note  the  phenomenon of wake-planet interaction 
discussed by \citet{Baruteau2013} that could lead to the same effect and was modelled through the introduction of orbital circularization. 

To illustrate how the results depend on the amount of circularization  (or wake-planet interaction) introduced
we have perfomed a simplified  illustrative  calculation by  extending
 the continuation of $1{\rm AU}\_1\_4\_32$ described above such that each run is extended for a further  $0.1t_{max},$ 
where $t_{max}$ is the duration of the run in the unextended survey,
during which  the circularization time for each planet taken to be $20$ orbits, as was adopted in \citet{Baruteau2013}. Results are shown in 
Fig. \ref{fig:1_4_longertimeF}. It will be seen that period ratios throughout the range $1.7-2$  are obtained.

The criteria  for  a run to have  terminated in  a second order resonance were applied  in the same way as for the other surveys 
and more are found after this doubly extended survey completes.
However, as in this case they are formed from divergent migration, capture into libration is not expected.
This is indeed found to be the case. Thus runs of the type illustrated in Fig. \ref{fig:resonances_example}, which end with resonant angles librating,
do not occur when the near second order  commensurabilities are obtained as a result of divergent migration produced through the action of efficient orbital
circularization.



\section{Summary and conclusions} \label{sec:conclusion}
In order to investigate the possible consequences for  potential  architectures of low mass planetary systems,
we  have performed numerical simulations of a pair of coplanar  migrating low-mass planets
in the mass range $ (1 - 4)\ \rmn{M}_{\oplus}.$
In  the  surveys we conducted, the planets were initialised on circular coplanar orbits with the outer one being at 1 AU or 5 AU.
Outside an inner cavity  of radius $0.08\ {\rm AU}$, the orbital evolution was considered for a  constant migration time, $\tau_{mig},$  in the range 
$(10^3 - 10^7)(M_\oplus/m_i) \ \rmn{yr}$  and constant ratio of circularization time to migration time  $\tau_c/\tau_{mig}$ in the range  $3\times 10^{-4} - 4.4\times 10^{-2}$.
These could be viewed as possibly arising from interaction with a protoplanetary disc.
The inner planet  was  initialised on an orbit  which was close to and exterior to either the  2:1 or the  3:2  mean motion resonance.
Note that  the length and time scales we used can be scaled to other values using the standard scalings for systems governed by gravity
(see Section \ref{initconfig}).

In order to provide a simple model of phenomena producing an inner cavity boundary to a protoplanetary disc, such as rapid changes in disc viscosity or interaction with the stellar magnetic field, we considered migration to terminate inside the cavity but performed 
some extended surveys for which  circularization was allowed  to continue for one or both planets.
We remark that the latter set up could be viewed as providing a heuristic modelling of  phenomena such as wake-planet interactions, or planet-density wave
interactions \citep[e.g.][]{Baruteau2013} or even residual planetesimal migration \citep[e.g.][]{Levison2014} that can halt or turn around convergent migration of the planets.
We remark that the work here is carried out in order to characterise the possible outcomes for migrating two planet systems
under very simple assumptions about forces leading to migration and circularization. It  does not constitute  population
synthesis studies.

Numerical simulations that terminated without an extended period of circularization resulted for the most part in,  either collisions between the planets
or, the formation of systems close to resonance. These were mostly first order with second order resonances occurring in a few $\%$ of cases.
Those were found to occur for modest eccentricities $< \sim 0.1$ and all resonant angles librated (see Section \ref{secondorderstudy}). 
This is in contrast to the situation where the near commensurability resulted from divergent migration of the type
encountered in the extended surveys. In those cases resonant trapping does not occur and the resonant angles circulate.
Thus the behaviour of the  resonant angles in systems near a second order commensurability  can indicate the form of any  migration that led to them.  

An analytic description of planets migrating in second order resonances was given in Sections \ref{sec:theory} - \ref{Behaviour}.
There we derived expressions for the ratio of the eccentricities of the planets and the absolute values of the eccentricities as a function
of the ratio of circularization time to migration time. The ratio of the eccentricities was subsequently evaluated correctly to fourth order in the eccentricities
for several cases and was found to be in reasonable agreement with simulations.
In a very small number of cases, period ratios corresponding to  even higher order resonances were seen. These were discussed in Sections \ref{Higherorder}
and \ref{Higherorderstudy}.

In order to obtain an extended  range of final  period ratios, it was necessary to extend the simulations so that the planet(s) underwent a  period of circularization without migration.   The important consequence of this is
that there is  a period  of effective divergent migration that causes the period  ratios to diverge. Depending on its effectiveness, the whole range of period ratios  in $(1, 2)$ could be obtained (see Section \ref{Continuationruns}). Thus, in order to determine final outcomes of planetary systems, the study of their interaction with the inner parts of a protoplanetary disc is crucial. A better understanding of the detailed inner disc structure, the effects of any density waves launched by an interaction with the central star and the interaction of any residual planetesimal disc will be required. These issues will form the focus of future studies.

\section*{Acknowledgments}

Xiang-Gruess acknowledges support through Leopoldina fellowship programme  (fellowship number LPDS 2009-50) and support of the Max Planck Society through a postdoctoral fellowship.
Simulations were performed using the Darwin Supercomputer of the University of Cambridge High Performance Computing Service, provided by Dell Inc. using Strategic Research Infrastructure Funding from the Higher Education Funding Council for England and funding from the Science and Technology Facilities Council.

\begin{appendix}

\section{The direct part of the disturbing function for second order resonances}\label{MD}
In order to discuss the $(p+2):p$  resonance,  the quantity $R_D$ is written in the compact general form
\begin{eqnarray}
&&R_D=\sum_{n,i }F_{i}^{n}\cos(n\theta_j  -i\varpi_1  +(i-2n)\varpi_2),
 \end{eqnarray}
where $\theta_j = (p+2)\lambda_2 - p\lambda_1.$ 
Here  the  index  of  summation  $n$  is  a non   negative integer  and  the  index of summation,  $i,$ is a positive or negative integer
or  zero. In addition terms that turn out to have the same cosine factor are collected together in the expressions
given below.
We take into account terms up to fourth order in the eccentricity  such that  $R_D$ takes  the explicit form
\begin{eqnarray}
&&\hspace{-7mm}R_D= F_{0}^0+F_{1}^0\cos(\varpi_2-\varpi_1)+F_{2}^0\cos2(\varpi_2 - \varpi_1)+  \nonumber\\
&&\hspace{-7mm}F_{2}^1\cos(\theta_j  -2\varpi_1)+ F_{1}^1\cos(\theta_j -\varpi_2-\varpi_1) + \nonumber  \\
&&\hspace{-7mm}F_{0}^1\cos(\theta_j -2\varpi_2)+F_{3}^1\cos(\theta_j+\varpi_2 -3\varpi_1)+
\nonumber  \\
&&\hspace{-7mm} F_{-1}^1\cos(\theta_j +\varpi_1-3\varpi_2)+F_{4}^2\cos(2\theta_j -4\varpi_1)+\nonumber \\
&&\hspace{-7mm} F_{3}^2\cos(2\theta_j-3\varpi_1-\varpi_2)+F_{2}^2\cos(2\theta_j-2\varpi_1 -2\varpi_2)+\nonumber \\
&&\hspace{-7mm} F_{1}^2\cos(2\theta_j -\varpi_1-3\varpi_2)+  F_{0}^2\cos(2\theta_j -4\varpi_2)\ ,
 \end{eqnarray}
where 
\begin{eqnarray}
&&\hspace{-7mm} F_{0}^0= f_0 + (e^{2}_1+e^{2}_2)f_2  + e_1^{4}f_{4} +  e_1^{2}e_2^{2} f_{5}+ e_2^{4}f_{6}  \nonumber\\
&&\hspace{-7mm}F_{1}^0= e_1e_2f_{10}+e^3_1e_2f_{11}+ e_1e_2^{3}f_{12}    \nonumber  \\
&&\hspace{-7mm}F_{2}^0= e_1^{2}e^{2}_2 f_{17} \nonumber  \\
&&\hspace{-7mm} F_{2}^1=  e^{2}_1f_{45} +  e^{2}_1e^{2}_2 f_{47}+  e_1^{4}f_{46}  \nonumber \\
&&\hspace{-7mm} F_{1}^1=  e_1e_2f_{49}+e_1^3e_2f_{50}+ e_1e_2^{3}f_{51}    \nonumber \\
&&\hspace{-7mm} F_{0}^1 = e_2^{2}f_{53} +  e_1^{2}e_2^{2} f_{54}+  e_2^{4}f_{55}\nonumber\\
&&\hspace{-7mm}F_{3}^1=  e_1^3e_2f_{68} \nonumber  \\
&&\hspace{-7mm}F_{-1}^1=   e_2^{3}e_1 f_{69}  \nonumber  \\
&&\hspace{-7mm} F_{4}^2=   e^4_1f_{90}\nonumber \\
&&\hspace{-7mm} F_{3}^2 =  e^3_1e_2f_{91}  \nonumber \\
&&\hspace{-7mm} F_{2}^2 =  e^{2}_1e_2^{2} f_{92} \nonumber \\
&&\hspace{-7mm} F_{1}^2= e_1 e_2^{3}f_{93}\nonumber \\
&&\hspace{-7mm} F_{0}^2= e_2^{4}f_{94} .
 \end{eqnarray}
the quantities $f_i$ and $f_{ij}$ are functions only of $\alpha= a_1/a_2$ and  their forms are tabulated in 
Appendix B of \citet{Murray1999}.

\end{appendix}

\label{lastpage}

\end{document}